\documentclass[12pt,a4paper]{article}
\usepackage[utf8x]{inputenc}
\usepackage[dvipsnames]{xcolor} 
\usepackage{jheppub}
\usepackage{rotating}
\usepackage{bm}        
\usepackage{amssymb}   
\usepackage{dsfont}
\usepackage{mathtools}
\usepackage{graphicx}
\graphicspath{ {./images/} }
\usepackage{tikz}
\usepackage{tikz-feynman}
\usepackage{mathrsfs}
\usepackage{braket}

\newcommand{\scri}{{\mathscr I}}

\definecolor{allOrderBlue}{rgb}{0.4,0.5,1}

\renewcommand{\[}{\begin{equation}\begin{aligned}}
\renewcommand{\]}{\end{aligned}\end{equation}}
\def\be{\begin{equation}}
\def\ee{\end{equation}}
\def\beal{\begin{equation}\begin{aligned}}
\def\eeal{\end{aligned}\end{equation}}
\def\nn{\nonumber}
\usetikzlibrary{patterns}

\renewcommand{\vec}[1]{\ensuremath{\boldsymbol{#1}}}

\newcommand{\xbar}{\eta}
\newcommand{\cA}{\mathcal{A}}

\newcommand{\rma}{{\rm a}}
\newcommand{\rmb}{{\rm b}}

\newcommand{\dd}{\mathrm{d}}

\newcommand{\sfk}{\mathsf{k}}
\newcommand{\sfh}{\mathsf{h}}

\newcommand{\vac}{\Omega}

\title{Amplitudes for Hawking Radiation }
\author[1]{Rafael Aoude,}
\emailAdd{rafael.aoude@ed.ac.uk}
\author[1, 2]{Donal O'Connell,}
\emailAdd{donal@ed.ac.uk}
\author[2]{Matteo Sergola}
\emailAdd{matteo.sergola@ipht.fr}

\affiliation[1]{Higgs Centre for Theoretical Physics,
School of Physics and Astronomy, \\
The University of Edinburgh, Edinburgh EH9 3JZ, Scotland, UK}
\affiliation[2]{
Institut de Physique Théorique, CEA, CNRS, \\Université Paris-Saclay, F–91191 Gif-sur-Yvette cedex, France}

\abstract{
We obtain the Hawking spectrum by exponentiating a series of Feynman diagrams describing a scalar field scattering through a collapse background.
Our approach is rooted in semiclassical methods of scattering amplitudes which have recently been developed for application to gravitational-wave physics.
The diagrams we encounter do not compute a standard amplitude, but rather an in-in generalisation of an amplitude which is closely connected to the Bogoliubov coefficients.
We also compute the subdominant one-loop correction in our perturbative approach, analogous to the triangle correction to Schwarzschild scattering. 
This term can be interpreted as a finite-size correction sensitive to the radius of the black hole.
}

\begin{document}
\maketitle

\section{Introduction}
\label{sec:Intro}

Scattering amplitudes have recently found a very successful new application to the study of gravitational wave physics.
This includes impressive computations of the interaction potential between two compact gravitating bodies, and of the (related) impulse during a scattering event.
The waveform itself is directly accessible to the methods of amplitudes during a scattering process.
Significant activity in this area has led to a number of interesting developments, recently including new insight on the significance of the choice of Bondi frame, for example.
See references~\cite{Arkani-Hamed:2017jhn,Bjerrum-Bohr:2018xdl,Cheung:2018wkq,Guevara:2018wpp,Chung:2018kqs,Kosower:2018adc,Bern:2019nnu,Bautista:2019tdr,Arkani-Hamed:2019ymq,Bjerrum-Bohr:2019kec,Bern:2020buy,Cristofoli:2020uzm,Monteiro:2020plf,Chung:2020rrz,Bjerrum-Bohr:2021vuf,Kalin:2019rwq,Damgaard:2021ipf,Herrmann:2021tct,Aoude:2021oqj,Bern:2021dqo,DiVecchia:2021bdo,Bern:2021yeh,Bautista:2021wfy,Damgaard:2019lfh,Aoude:2020onz,
Cristofoli:2021vyo,Herrmann:2021lqe,Bjerrum-Bohr:2022blt,Bautista:2022wjf,Bern:2023ccb, Bellazzini:2022wzv,Bern:2024adl,Driesse:2024xad,Gambino:2024uge,Luna:2023uwd,Elkhidir:2023dco,Herderschee:2023fxh,Brandhuber:2023hhy,Brunello:2024ibk,Cristofoli:2021jas,Mogull:2020sak,Jakobsen:2021smu,Jakobsen:2021lvp,Bini:2024rsy,Alaverdian:2024spu,Aoki:2024boe,Luna:2017dtq,Cangemi:2022bew,  Barack:2023oqp, Georgoudis:2023lgf,Dlapa:2021npj,Bini:2023fiz,DeAngelis:2023lvf,Brandhuber:2023hhl,Aoude:2023dui,Georgoudis:2023eke,Bohnenblust:2023qmy,Falkowski:2024bgb,Amalberti:2024jaa,Porto:2024cwd,Kalin:2022hph,Dlapa:2024cje,Haddad:2021znf,AccettulliHuber:2020oou,Gatica:2023iws,Caron-Huot:2023ikn,Adamo:2023cfp,Adamo:2017nia,Damgaard:2024fqj,Mougiakakos:2024lif,Mougiakakos:2024nku,Bautista:2024emt,Bautista:2024agp,Haddad:2024ebn,Aoude:2024jxd,Driesse:2024feo,Bohnenblust:2024hkw,Cangemi:2023bpe} for a sampling of the literature, and~\cite{Travaglini:2022uwo,Kosower:2022yvp,Buonanno:2022pgc} for recent reviews.

All of this work obviously relies on progress in our understanding of how to extract classical physics from quantum field theory.
But of course there are many interesting and important questions in \emph{semiclassical} physics.
This raises an important question: can we also apply the new methods of amplitudes to determine the results of semiclassical observables?

In this paper we answer this question affirmatively in the case of Hawking radiation.
In fact, Hawking's original calculation is remarkably close in spirit to the current programme of extracting classical observables from amplitudes.
As we review below in section~\ref{sec:Review}, the key step in Hawking's paper is to determine what is essentially a scattering amplitude in a situation where a classical wave propagates through a ``collapse background'': a time-dependent background describing the formation of a black hole.

We will recover this amplitude in section~\ref{sec:Tree_level} using the (semi-)classical methods of scattering amplitudes, closely following the KMOC setup~\cite{Kosower:2018adc}.
As is familiar from our understanding of the amplitude-action relation~\cite{Bern:2021dqo} and the eikonal formalism, dominant classical tree diagrams exponentiate into a phase.
This exponentiation, which we demonstrate explicitly, is the key result of our paper.
The steps are in fact very familiar because Hawking's scattering process involves a high frequency wave scattering from a black hole as it forms.
The large frequency plays a role entirely analogous to the small $\hbar$ approximation familiar from the classical limit of amplitudes.

Of course Hawking radiation and its ramifications are a major area of study in theoretical physics.
The literature is vast; for a review of the whole field, we can recommend the lectures~\cite{Harlow:2014yka} and the review~\cite{Page:2004xp}. 
Our work is far from the first to combine ideas from gravitational waves with aspects of Hawking radiation: we can point, for example, to references~\cite{Goldberger:2020wbx,Goldberger:2020geb,Kim:2020dif,Ilderton:2023ifn,Gaddam:2021zka,Gaddam:2020mwe,Ferreira:2020whz,Melville:2023kgd}.

Hawking deduced the thermal spectrum of emitted radiation using the theory of Bogoliubov coefficients, which allows for an analytic continuation from a $1 \rightarrow 1$ scattering process to a $0 \rightarrow 2$ pair-production process.
We argue that the Bogoliubov coefficients are themselves familiar generalisations of amplitudes in section~\ref{sec:Bogoliubov}.
The generalised amplitudes are of the type introduced recently by Caron-Huot, Giroux, Hannesd\'ottir and Mizera~\cite{Caron-Huot:2023vxl,Caron-Huot:2023ikn} in their study of crossing, and were also studied by Schwarz~\cite{Schwarz:2019ggp,Schwarz:2019npn}.
They involve more complicated boundary conditions than the standard in/out boundary conditions.
The waveform $\braket{0 | a(p_1') a(p_2') S^\dagger a(k) S a^\dagger(p_1) a^\dagger(p_2)|0}$ is an in-in example.
In the case of Hawking radiation, the generalised amplitudes describe quantum-mechanical particles scattering on a fixed classical background (rather than vacuum scattering as in references~\cite{Caron-Huot:2023vxl,Caron-Huot:2023ikn}.)
However amplitudes on backgrounds are also an important component of the amplitudes/gravitational waves programme~\cite{Adamo:2021rfq,Menezes:2022tcs,Adamo:2022qci,Adamo:2022rmp,Cristofoli:2022phh,Adamo:2023cfp} so this aspect of Hawking's approach is also familiar.

Our work shares many features of the eikonal approach to classical scattering, which has been particularly emphasised for instance in references~\cite{DiVecchia:2020ymx,DiVecchia:2021bdo,DiVecchia:2022nna,DiVecchia:2022piu,Georgoudis:2023eke,Georgoudis:2024pdz,Heissenberg:2024umh}, see~\cite{DiVecchia:2023frv} for a very comprehensive review.
As in the case of eikonal exponentiation, the phase in Hawking's scattering computation itself has a perturbative expansion.
Exponentiated tree diagrams recover the leading term in the phase in both cases.
In section~\ref{sec:OneLoop} we show that subdominant one-loop corrections compute the first perturbative correction in the expansion of Hawking's phase.
This requires us to separate terms which exponentiate into the phase from other terms which form part of a ``remainder'', analogous to the quantum remainder in the eikonal story.
Physically, the origin of the perturbative expansion of Hawking's phase is the finite radius $2 GM$ of the event horizon; our perturbative expansion is in powers of this distance over the closest approach of our classical wave from the horizon. 
As a result, the correction we compute in section~\ref{sec:OneLoop} can be interpreted as a finite-size correction to the dynamics.
Remarkably these correction terms are irrelevant for determining the thermal distribution of Hawking radiation.
We conclude our work in section~\ref{sec:Conclusions} with a discussion and some comments on future work.

\subsection*{Conventions}{ We work in four dimensions with the mostly negative signature $(+,-,-,-)$. Factors of $2\pi$ are absorbed as in \cite{Kosower:2018adc} $\hat{\dd}^np = \dd^n p /(2\pi)^n$ and $\hat{\delta}^n(\cdot) = (2\pi)^n \delta^n(\cdot)$. We set $c=1$ but keep $G$ explicit. Unless specified, limits of the integral are from $-\infty$ to $+\infty$. We use the notation $\sfk^\mu \equiv \sfk^\mu(x)$ for the null vector appearing in the Kerr-Schild metric, while $k^\mu$ instead denotes a momentum vector.}

\section{Hawking's scattering computation}
\label{sec:Review}

Let us begin by reviewing Hawking's original calculation~\cite{Hawking:1975vcx}, emphasising the on-shell quantity, essentially an amplitude, which plays a central role in his work.
We study a particle scattering through a spacetime as matter collapses into a black hole.
It is enough to treat the collapse as a gravitational background, so we may write the spacetime metric as
\[\label{eq:metricvaid}
\dd s^2 = \dd s_0^2 - \frac{2 GM(v)}r \dd v^2 \,,
\]
where $\dd s_0^2$ is a flat background metric in spherical coordinates
\[
\dd s_0^2 = \dd t^2 - \dd r^2 - r^2 \dd \Omega^2 \,,
\]
and $v = t + r$ is the advanced (Kerr--Schild) time. The function $M(v)$ is an increasing function of $v$, which we take to be zero in the past and $M$ in the future.
This background is known as the Vaidya metric~\cite{Vaidya:1966zza,Stephani_Kramer_MacCallum_Hoenselaers_Herlt_2003}.
It is a Kerr--Schild (KS) solution of the Einstein equations, though not a vacuum solution: a stress tensor
\[
T^{\mu\nu}(x)=\frac{M' (v)}{4\pi \, r^2} \sfk^\mu(x)\sfk^\nu(x) 
\]
is present describing matter collapsing into the black hole.
As the Hawking process is universal, we may simplify our discussion by taking the mass function\footnote{Note that the Schwarzschild black hole case is obtained  by setting $M(v)=M$ to be constant.} to be 
\[\label{massv}
M(v) = M \, \Theta(v) \,.
\]

In what follows, it will often be convenient to refer to the vector and tensor fields that characterise the line element \eqref{eq:metricvaid}. It is useful to write the metric as
\begin{equation}\label{metric}
    g_{\mu\nu}(x)=\eta_{\mu\nu}-\frac{2 G M(v)}{r}\sfk_\mu(x)\sfk_\nu(x),\,\,\,\,\,\, \sfk_\mu(x)=\left(1, \frac{\vec{x}}{r}\right),
\end{equation}
with $r=|\vec{x}|$. 
The $\sfk^\mu(x)$ vector field has the crucial property that it squares to zero
\begin{equation}\label{squaredd}
    \sfk_\mu(x)\sfk^\mu(x)=0,
\end{equation} 
which implies that the inverse metric is simply 
\begin{equation} \label{eq:invMetric}
    g^{\mu\nu}(x)=\eta^{\mu\nu}+\frac{2 G M(v)}{r}\sfk^\mu(x)\sfk^\nu(x).
\end{equation}
It also follows that $\sqrt {|g|} =1$.
For later convenience, it will also be useful to define the quantity 
\begin{equation}\label{metrich}
     h_{\mu\nu}(x)=\frac{2GM(v)}{r}\sfk_\mu (x)\sfk_\nu(x), 
\end{equation}
which plays a similar role to the metric perturbation. However,  it is important to stress that the expression \eqref{metric} is \emph{exact} in the background to all orders in $G$; equation~\eqref{metrich} involves no perturbative expansion despite appearances.

Following Hawking, we consider a high-frequency solution of the Klein-Gordon equation of the form
\[\label{eq:highFreq}
\phi(x) = \exp \big( i s(x) / \xbar \big) \,,
\]
where $\xbar$ is a small number so that the quantity $s(x)/\xbar$ is large, corresponding to the high frequency approximation. 
(Later in this article, small $\eta$ will play a role entirely analogous to small $\hbar$ in classical applications of amplitudes.)
To leading order in $\xbar$, the Klein-Gordon equation implies that the phase function $s(x)$ satisfies
\[
\label{eq:HamiltonJacobi}
g^{\mu\nu} (\partial_\mu s)(\partial_\nu s) = 0 \,.
\]
This is the Hamilton-Jacobi equation: the assumption of large frequency leads to the geometric optics approximation.

In order to relate particles in the future to particles in the past, we need to find a way to relate the phase function in the future (on $\scri^+$) to the phase in the past (on $\scri^-$).
An important consequence of the Hamilton-Jacobi equation~\eqref{eq:HamiltonJacobi} is that the vector field $\partial_\mu s$ is null and geodesic.
In other words, the normal vectors to the phase function are tangent vectors of lightlike rays.
We will specialise to the situation where these rays are radial to exploit the spherical symmetry of the Vaidya metric.

Hawking used his geometric knowledge of null geodesics in Schwarzschild to relate the phase in the future to the phase in the past in quite a general collapse scenario. 
Here we follow a more pedestrian route, specialising to the Vaidya case.

\begin{figure}[!htb]
\centering
\usetikzlibrary{decorations.markings,decorations.pathmorphing}
\usetikzlibrary{angles,quotes} 
\usetikzlibrary{arrows.meta} 

\newcommand{\calI}{\mathscr{I}} 
\tikzset{>=latex} 
\colorlet{myred}{red!80!black}
\colorlet{myblue}{blue!80!black}
\colorlet{mygreen}{green!80!black}
\colorlet{mydarkred}{red!50!black}
\colorlet{mydarkblue}{blue!50!black}
\colorlet{mylightblue}{mydarkblue!3}
\colorlet{mypurple}{blue!40!red!80!black}
\colorlet{mydarkpurple}{blue!40!red!50!black}
\colorlet{mylightpurple}{mydarkpurple!80!red!6}
\colorlet{myorange}{orange!40!yellow!95!black}
\colorlet{shadecolor}{gray!40}
\definecolor{darktangerine}{rgb}{1.0, 0.66, 0.07}
\tikzstyle{cone}=[mydarkblue,line width=0.2,top color=blue!60!black!30,
                  bottom color=blue!60!black!50!red!30,shading angle=60,fill opacity=0.9]
\tikzstyle{cone back}=[mydarkblue,line width=0.1,dash pattern=on 1pt off 1pt]
\tikzstyle{world line}=[myblue!60,line width=0.4]
\tikzstyle{world line t}=[mypurple!60,line width=0.4]
\tikzstyle{particle}=[mygreen,line width=0.5]

\tikzstyle{singularity}=[mydarkblue,line width=0.7,decorate,decoration={snake,amplitude=0.9,segment length=4,post length=3.8}]

\tikzstyle{photon}=[-{Latex[length=4,width=3]},myorange,line width=0.6,decorate,
                    decoration={snake,amplitude=0.9,segment length=4,post length=3.8}]


\tikzset{declare function={%
  penrose(\x,\c)  = {\fpeval{2/pi*atan( (sqrt((1+tan(\x)^2)^2+4*\c*\c*tan(\x)^2)-1-tan(\x)^2) /(2*\c*tan(\x)^2) )}};%
  penroseu(\x,\t) = {\fpeval{atan(\x+\t)/pi+atan(\x-\t)/pi}};%
  penrosev(\x,\t) = {\fpeval{atan(\x+\t)/pi-atan(\x-\t)/pi}};%
  kruskal(\x,\c)  = {\fpeval{asin( \c*sin(2*\x) )*2/pi}};
}}
\def\tick#1#2{\draw[thick] (#1) ++ (#2:0.04) --++ (#2-180:0.08)}
\def\Nsamples{40} 

\def\R{0.08} 
\def\e{0.08} 
\def\ang{45} 
\def\angb{acos(sqrt(\e)*sin(\ang))} 
\def\a{\R*sin(\ang)*sqrt(1-\e*sin(\ang)^2)/(1-\e*sin(\ang)^2)} 
\def\b{\R*sqrt(\e)*sin(\ang)*cos(\ang)/(1-\e*sin(\ang)^2)} 
\def\coneback#1{ 
  \draw[cone back] 
    (#1)++(-45:\R) arc({90-\angb}:{90+\angb}:{\a} and {\b});
  \draw[cone,shading angle=-60] 
    (#1)++(0,{\R*cos(\ang)/(1-\e*sin(\ang)^2)}) ellipse({\a} and {\b});
}
\def\conefront#1{ 
  \draw[cone] 
    (#1) --++ (45:\R) arc({\angb-90}:{-90-\angb}:{\a} and {\b})
     --++ (-45:2*\R) arc({90-\angb}:{-270+\angb}:{\a} and {\b}) -- cycle;
}

\begin{tikzpicture}[scale=3.5]
  \message{Penrose diagram (radius r)^^J}
  
  \def\Nlines{4} 
  \def\ta{tan(90*1.0/(\Nlines+1))} 
  \def\tb{tan(90*2.0/(\Nlines+1))} 
  \coordinate (O) at ( 0, 0); 
  \coordinate (S) at ( 0,-1); 
  \coordinate (N) at ( 0, 1); 
  \coordinate (E) at ( 1, 0); 
  \coordinate (BH1) at (0, 0.5);  
  \coordinate (BH2) at (0.5, 0.5);  
  \coordinate (X) at ({penroseu(\tb,\tb)},{penrosev(\tb,\tb)});
  \coordinate (X0) at ({penroseu(\ta,-\tb)},{penrosev(\ta,-\tb)});

  \fill[mylightblue] (BH1) -- (BH2) --(E) -- (S) -- cycle;

  \foreach \i [evaluate={\c=\i/(\Nlines+1); \ct=tan(90*\c);}] in {1,...,\Nlines}{
    \message{  Running i/N=\i/\Nlines, c=\c, tan(90*\c)=\ct...^^J}
    \draw[world line,shadecolor,samples=\Nsamples,smooth,variable=\r,domain=-1:0.5] 
      plot({penrose(\r*pi/2,\ct)},\r);
  }
  
  \draw[thick,mydarkblue] (BH2) -- (E) -- (S) -- (BH1) ;
  \draw[thick,mylightblue] (BH1) -- (BH2) -- cycle;
  \draw[-,singularity] (0,0.5) -- (0.5,0.5) node[above=2,right=1] {};

  \node[mydarkblue,above right,align=right] at (-53:0.77) {$v_0$};
  \draw[thick,mydarkblue] (0.5,-0.5) -- (0.0,0.0);
  \draw[thick,mydarkblue] (BH2) -- (0.0,0.0);

  \draw[->,photon] (0.40,-0.60) -- (0.3,-0.50) node[above=2,left=1] {};
  \draw[solid,thick,gray] (0.3,-0.50) -- (0.0,-0.20);
  \draw[solid,thick,gray] (0.0,-0.20) -- (0.50,0.30); 
  \draw[->,photon,myred]  (0.50,0.30) -- (0.60, 0.4) node[above=2,left=1] {};
  \node[mydarkblue,above right,align=right] at (0.38,-0.7) {$v$};

  \node[mydarkblue,above right,align=right] at (+10:0.85) {$\calI^+$};
  \node[mydarkblue,below right,align=right] at (-10:0.80) {$\calI^-$};

  \draw[->,photon] (0.60,-0.40) -- (0.50,-0.30) node[above=2,left=1] {};
  \draw[dashed,thin,gray] (0.0,0.20) -- (0.50,-0.30); 
  \node[mydarkblue,above right,align=right] at (0.575,-0.50){$v{=}0$};
  

  \draw[-,thick,black] (0.1,-0.30) -- (0.20,-0.20); 
  \node[mydarkblue,above right,align=right] at (0.130,-0.345) {$ \varepsilon$};
    
  
  
\end{tikzpicture}
\caption{Penrose diagram describing the redshift induced by a gravitational collapse. Some important points in this diagram are at $v=0$ (describing the lightlike infalling  radiation)  and at $v_0=-4GM$ (describing the event horizon).}
\label{fig:penrose}
\end{figure}
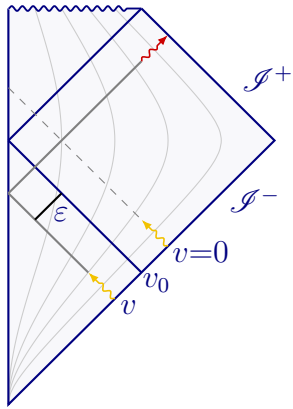

Consider an incoming null geodesic ``starting'' on $\scri^-$. 
Because of spherical symmetry, the geodesic is characterised only by the advanced time coordinate $v$ where it ``meets'' $\scri^-$, see figure~\ref{fig:penrose}.
We assume that $v < 0$.
It is useful to distinguish between the value of $v$ on $\scri^-$ corresponding to our ray and other values of $v$ in the bulk of spacetime; we therefore use an upper-case notation $V$ for values of the advanced time away from $\scri^-$ in this discussion.

The geodesic is initially in the flat region, following the line $V=v$ to the origin.
At the origin $r = 0$ we are still in the flat $V <0$ region, so the geodesic transitions to an outgoing ray of constant retarded time $U \equiv V-2r$.
As the retarded time equals the advanced time at the origin of radial coordinates, the constant $U$ on this part of the geodesic equals the initial value of $V$ on $\scri^-$, namely $U=v$.

The situation becomes more interesting when the geodesic transitions into the curved region $V > 0$.
We assume that at $V = 0$, the ray is just outside the event horizon of the black hole. 
This horizon is at $r = 2GM$, so we require that our geodesic has $V = 0$ at $r = 2GM + \varepsilon$ for small positive $\varepsilon$.
We further assume that the geodesic is continuous\footnote{This assumption can be justified as follows. By increasing the total mass $M$ in the infalling radiation, we can make the event horizon as large as we wish, and therefore we can choose the density of matter to be arbitrarily low when our probe particle passes through it. By taking this density very low, it is clear that the (high energy) probe particle passes through the infalling matter without suffering any large deflection. Hence its geodesic is continuous when it meets the matter.} at $V = 0$. 
Then at this point we know that $v = U = -2r = -2(2 GM + \varepsilon)$, so
\[\label{eq:epsilon}
v = - 4 GM - 2\varepsilon \,.
\]
The ray which coincides with the horizon has $\varepsilon = 0$, so its initial value of advanced time is
$v_0 = -4 GM$, allowing us to eliminate $\varepsilon$ in favour of $v_0$:
\[
\varepsilon = \frac12 \left( v_0 - v \right) \,.
\]

In the curved region, the outgoing geodesic satisfies
\[
\dd s^2 = 0 = \left(1 - \frac{2GM}{r} \right) \dd V^2 - 2 \dd V \, \dd r\, .
\]
As $\dd V \neq 0$, the function $V(r)$ therefore satisfies a simple ordinary differential equation.
Integrating, and imposing continuity at $V = 0$, we have
\[\label{eq:CoulombPhase}
V -2 r - 4GM \log \left( \frac{2 r + v_0}{\mu} \right) = v - 4 GM \log \left( \frac{v_0 - v}{\mu} \right) \,.
\]
The quantity $\mu$ can be chosen freely and will drop out of the computation below.
Notice that the logarithmic dependence on radius on the left-hand side is the usual Coulomb phase.
We deal with this (again, following the literature~\cite{Hawking:1975vcx}) by absorbing the logarithm into the definition of retarded time on $\scri^+$.
The conclusion is that geodesics which are characterised by advanced time $v < 0$ at $\scri^-$ are mapped to a constant retarded time on $\scri^+$ given by 
\[\label{eq:rayTracingRelation}
u &\equiv V -2 r - 4GM \log (2 r +v_0) / \mu \\
&= v - 4 GM \log (v_0 - v)/\mu \,.
\]
We will refer to this important result as the ray-tracing relation.
It relates rays which ``start'' at $v <0$ on $\scri^-$ to their endpoint $u$ on $\scri^+$.  
On the other hand, geodesics which start at $v \geq 0$ do not reach $\scri^+$ --- they either go into the black hole, or form its event horizon.

We can use the ray-tracing relation to follow the propagation of a state (in the geometric optics approximation). 
The wavefunction of a spherically symmetric energy eigenstate at $\scri^+$, up to a phase which we discuss in more detail in appendix~\ref{app:sphericalNorms}, is
\[
F(u, E) = \frac{\exp(-i E u)}{2 E r} \,.
\]
We have chosen the notation $F$ for this function because it is a Future single-particle energy eigenstate at $\scri^+$, namely the wavefunction of an out state.
Propagating the wavefunction of this state back to $\scri^-$ using the relation~\eqref{eq:rayTracingRelation}, we conclude that (in the geometric optics approximation) the wavefunction of this state is 
\[
F(v, E) &= 
\begin{cases}
     \displaystyle{\frac{1}{2 Er}} \exp\biggl[-i E v + 4 i \, GM \, E \, \log \Bigl((v_0 - v)/\mu\Bigr) \biggr]\,, \qquad &v < v_0 \,, \\
    0 \,, &v > v_0 \,.
\end{cases}
\]
on $\scri^-$. 
The wavefunction of the out state is non-trivial on $\scri^-$, because there is a non-trivial $S$ matrix.
On the other hand, the wavefunction of a Past spherically-symmetric single-particle state of energy $E_0$  on $\scri^-$ is trivial:
\[\label{eq:incomingWavefunction}
P(v, E_0) = \frac{\exp(-i E_0v)}{2 E_0 r} \,.
\]
The ``Hawking amplitude'', thought of as the overlap between the incoming and outgoing wavefunctions, is then
defined to be
\[\label{eq:Ahawkfull}
\mathcal{A}_H &= \int_{-\infty}^\infty \dd v \, (8 \pi r^2) \, F^*(v, E) \, i \partial_v P(v, E_0) \\
&= 
\frac{2 \pi}{E} \int_{-\infty}^{v_0} \dd v \, 
\exp \left[-4 i GM \, E \log \big( (v_0 - v) / \mu \big) + i (E - E_0) v\right] \,.
\]
The factor $8 \pi r^2$ appearing here is a normalisation; the $v$ derivative supplies a factor of energy required for an overlap of states in relativistic quantum mechanics.
Otherwise this expression is simply the overlap of wavefunctions familiar from undergraduate quantum mechanics.
However, the object $\mathcal{A}_H$ is not quite an amplitude, but it is a member of a class of on-shell objects which generalises amplitudes.
We will discuss these generalised amplitudes in more detail in section~\ref{sec:Bogoliubov} below.
For now we comment that $\mathcal{A}_H$ is analogous to an amplitude, and contains all the information about the thermal distribution of Hawking radiation. 

The presence of the black hole's horizon leads to a dichotomy in the space of possible future states. 
In addition to the usual modes on $\scri^+$, there are also modes associated with the horizon itself.
A generic out state may be written in the form
\[
\begin{pmatrix}
    \ket{\textrm{out}, \scri} \\
    \ket{\textrm{out}, \mathcal{H}}
\end{pmatrix}
\]
corresponding to an out state on $\scri^+$ and on the horizon $\mathcal{H}$.
The (generalised) amplitude in equation~\eqref{eq:Ahawkfull} is an overlap only with respect to part of this full Hilbert space.
We can therefore view equation~\eqref{eq:Ahawkfull} as a ``reduced'' (generalised) amplitude, in analogy with the process of performing a partial trace over a subspace of the Hilbert space at the level of density matrices.
We will discuss this point further in section~\eqref{spectrum}.

It is very natural to study this Hawking amplitude given recent progress in our understanding of scattering amplitudes, and their generalisations, in gravity (and beyond).
We will reconstruct it using the modern semiclassical machinery of scattering amplitudes, which we can directly apply to the geometric optics approximation by formally identifying the parameter $\xbar$ in equation~\eqref{eq:highFreq} with $\hbar$.
Our computation will be perturbative in the exponent, expanding in $v / v_0$ and working at next-to-leading order.
For that reason, we expect to encounter the amplitude in the form
\[
\label{eq:AHawking}
\mathcal{A}_H &\simeq
\frac{2\pi}{E} \int_{-\infty}^{v_0} \dd v \,
\exp\biggl[-4 i \, GM \, E \, \log (- v/\mu) - {16 i \, G^2M^2 \, E}/{v} + i (E-E_0) v \biggr]  \,.
\]

\section{Resummation of Feynman diagrams}\label{sec:Tree_level}

The original derivation of Hawking radiation~\cite{Hawking:1975vcx} and the seminal work of Wald~\cite{Wald:1975kc,Wald:1995yp} used the machinery of Bogoliubov coefficients combined with a suitable version of the ray-tracing relation.
We follow a different approach, instead identifying and resumming a class of Feynman diagrams. 
This resummation accounts for the non-trivial ray trajectories in Hawking's original argument, and is valid in the geometric optics limit.
Before describing the diagrams, let us discuss the relevant scales in our setup.

We scatter a massless scalar particle off the Vaidya background with impact parameter $b^\mu$. 
This particle has wavelength $\lambda \sim \hbar / E = \hbar / |\mathbf{p}|$; in the geometric optics limit we must require as $|\mathbf{b}|\gg \lambda$ (see the discussion in~\cite{Cristofoli:2021vyo} for more details). 
If we relate the impact parameter with the momentum transfer by $|\mathbf{b}| = {\hbar}/{|\mathbf{q}|}$, we can expand our amplitudes using the inequalities
\[\label{eq:geolocks}
\textrm{geometric-optics limit:} \qquad 
|\mathbf{b}|\gg \lambda,
\qquad
|\mathbf{q}|\ll |\mathbf{p}| 
\qquad
\textrm{or}
\qquad
\eta \equiv |\mathbf{q}|/|\mathbf{p}| \ll 1.
\]
Note that this is slightly different from the usual classical expansion since our incoming particle $p$ is massless and in principle should also scale with $\hbar$. 
However in geometric optics, this is a hard particle, so it is treated analogously to a massive particle in the classical expansion.
In the following, we will expand in ``orders'' of the geometric optics limit, keeping the leading order terms in $\eta$ but still expanding in the coupling parameter $(GM)$. This will allow us to use the fragment notation of~\cite{Cristofoli:2021jas}, albeit replacing the classical order by the geometric-optics order.
With these minor tweaks, we can use the customary machinery of semi-classical scattering amplitudes.

In this section we follow the usual conventions of scattering amplitudes.
In particular, we will evolve an initial state forwards in time, in distinction from the ray-tracing discussion in the previous section which propagated a future state backwards in time.
We will compare forwards and backwards time evolution later in section~\ref{sec:forwardsVbackwards}.

\subsection{Tree-level Vaidya scattering} 

We are now ready to scatter our particle off the Vaidya background, this time using the machinery of amplitudes.
We follow especially references~\cite{Kosower:2018adc,Cristofoli:2021jas} and
start by specifying the initial KMOC state describing the incoming massless scalar
\begin{align}\label{state}
    \ket{\psi} = \int\! 
    \dd\Phi(p)\varphi(p)|p\rangle 
     = \int\! 
    \dd\Phi(p)\ket{p}\int \! \dd v \, e^{iEv}
    {\varphi}(v) \,,
\end{align}
where
\[
\dd \Phi(p) = \hat{\dd}^4 p \, \hat{\delta}(p^2) \Theta(p^0) 
\]
is the on-shell phase space measure, and a momentum eigenstate is defined to be
\begin{equation}
    \ket{p}= a^\dagger (p)\ket{\vac  } \,.
\end{equation}
Here, the state $\ket{\vac}$ is the vacuum of the scalar theory in the non-trivial Vaidya background.
Note that above we Fourier transformed the wavefunction to position space as follows
\begin{equation}
    \varphi(p)=\int \dd v\, e^{i Ev}{\varphi}(v),
\end{equation}
which will be very convenient later. 
In our notation,  $E=|\vec{p}|$, and so
the wavefunction $\varphi(p)$ depends on one real variable $\varphi(p)=\varphi(E)$.
Consequently the state $\ket{\psi}$ is spherically symmetric: the integral over the angles of the momentum $p$ in equation~\eqref{state} is an angular average.
Notice that the phase $e^{iEv}$ in this state plays a role analogous to the translation $e^{ib\cdot p}$ which is normally inserted in a KMOC state. 
However, since in the present case the translation is purely timelike, it seems convenient to \emph{define} an ``impact'' parameter 
\begin{equation}
    b^\mu(v)=(v,\vec{0}).
\end{equation}
This has the physical interpretation of the time at which the probe particle passes through the origin before the infalling matter reaches the origin. 
Additionally, we will see that this  definition of the impact parameter makes the connection with post-Minkowskian amplitude methods more explicit.

We now evolve the state in time using the $S$-matrix,  and project onto a single particle state of momentum $p'$:
\[
\label{sstate}
 \braket{p'|S|\psi}&= \int \dd\Phi(p) \int \! \dd v \, e^{ip\cdot b(v)}
    \varphi(v) \,\langle p'| S|p\rangle \,.
\]
To simplify equation~\eqref{sstate} further, we change variable of integration from $p$ to $q \equiv p'-p$, which is useful in view of the geometric-optics hierarchy of momenta~\eqref{eq:geolocks}. 
Focussing on the interaction part only, we find
\begin{equation}\label{eq:fullInteraction}  
 \braket{p'|S-1|\psi}=\int \dd v \,\varphi(v) e^{i p'\cdot b(v)}\int \hat{\dd}^4 q\, \hat{\delta}(2 p'\cdot q+q^2)i\mathcal{A} (p'-q\to p')e^{-iq\cdot b(v)}.
\end{equation}

To model the probe limit dynamics of the scalar it is sufficient  to specify the following action:
\begin{equation}
   S_{\rm action}=\frac{1}{2}\int \dd^4 x\, \sqrt{|g|} g^{\mu\nu}\partial_\mu \phi\, \partial_\nu \phi \,,
\end{equation} 
where the metric $g_{\mu\nu}$ is the Vaidya metric~\eqref{metric}. 
As before,  we will  employ Kerr--Schild coordinates for the background; this drastically simplifies the perturbation theory.\footnote{For related applications of Kerr-Schild solutions and their relation to amplitudes we refer to  \cite{Monteiro:2014cda, Luna:2015paa,Luna:2016due,Monteiro:2020plf,  Menezes:2022tcs,Carrillo-Gonzalez:2017iyj,Bahjat-Abbas:2017htu,Vines:2017hyw,Bianchi:2023lrg,Berman:2018hwd,Ridgway:2015fdl,White:2020sfn,Godazgar:2020zbv,Monteiro:2021ztt,Stephani_Kramer_MacCallum_Hoenselaers_Herlt_2003}.} 
In fact, using equation~\eqref{eq:invMetric} and the usual properties of Kerr--Schild metrics, the action simplifies to
\begin{equation}
     S_{\rm action}=   S_{\rm free} +\frac{1}{2}\int \dd^4 x\, \frac{2GM(t+r)}{r}\left(\partial_t\phi-\partial_r\phi\right)^2,
\end{equation}
with $\sfk^\mu \partial_\mu=\partial_t-\partial_r$.
As we see KS coordinates are an extremely convenient choice as the interaction is simple and is linear in $GM$. This is to be contrasted  (for instance) with harmonic coordinates; there, metric non-linearities arise at each new order in $GM$ from expanding $\sqrt{|g|}, \,\,g^{\mu\nu}$ in the action. 

We can now extract the momentum space amplitude by standard techniques. 
The interaction can be represented graphically as a source injecting momentum $q$ into the worldline of the scalar particle

\[
\begin{tikzpicture}[x=0.75pt,y=0.75pt,yscale=-1,xscale=1]

  
\tikzset {_i1th39cse/.code = {\pgfsetadditionalshadetransform{ \pgftransformshift{\pgfpoint{0 bp } { 0 bp }  }  \pgftransformrotate{0 }  \pgftransformscale{2 }  }}}
\pgfdeclarehorizontalshading{_ehsthm2yr}{150bp}{rgb(0bp)=(1,1,1);
rgb(42.606724330357146bp)=(1,1,1);
rgb(45.89285714285714bp)=(0.82,0.89,0.97);
rgb(57.14285714285714bp)=(0.29,0.56,0.89);
rgb(57.25bp)=(0.29,0.56,0.89);
rgb(62.5bp)=(0.29,0.56,0.89);
rgb(100bp)=(0.29,0.56,0.89)}

  
\tikzset {_ssoze1dis/.code = {\pgfsetadditionalshadetransform{ \pgftransformshift{\pgfpoint{0 bp } { 0 bp }  }  \pgftransformrotate{0 }  \pgftransformscale{2 }  }}}
\pgfdeclarehorizontalshading{_02783xg5u}{150bp}{rgb(0bp)=(1,1,1);
rgb(42.428152901785715bp)=(1,1,1);
rgb(47.67857142857143bp)=(0.82,0.89,0.97);
rgb(52.23214285714286bp)=(0.29,0.56,0.89);
rgb(57.5bp)=(0.29,0.47,0.89);
rgb(62.42815290178571bp)=(0.29,0.47,0.89);
rgb(100bp)=(0.29,0.47,0.89)}
\tikzset{every picture/.style={line width=0.75pt}} 


\draw    (332.18,16451.69) -- (286,16443) ;
\draw [shift={(309.09,16447.35)}, rotate = 190.66] [fill={rgb, 255:red, 0; green, 0; blue, 0 }  ][line width=0.08]  [draw opacity=0] (5.36,-2.57) -- (0,0) -- (5.36,2.57) -- cycle    ;
\draw    (378,16444) -- (332.18,16451.69) ;
\draw [shift={(355.09,16447.85)}, rotate = 170.47] [fill={rgb, 255:red, 0; green, 0; blue, 0 }  ][line width=0.08]  [draw opacity=0] (5.36,-2.57) -- (0,0) -- (5.36,2.57) -- cycle    ;
\draw [color={rgb, 255:red, 0; green, 0; blue, 0 }  ,draw opacity=1 ]   (333.68,16451.69) .. controls (335.35,16453.36) and (335.35,16455.02) .. (333.68,16456.69) .. controls (332.01,16458.36) and (332.01,16460.02) .. (333.68,16461.69) .. controls (335.35,16463.36) and (335.35,16465.02) .. (333.68,16466.69) .. controls (332.01,16468.36) and (332.01,16470.02) .. (333.68,16471.69) .. controls (335.35,16473.36) and (335.35,16475.02) .. (333.68,16476.69) .. controls (332.01,16478.36) and (332.01,16480.02) .. (333.68,16481.69) .. controls (335.35,16483.36) and (335.35,16485.02) .. (333.68,16486.69) .. controls (332.01,16488.36) and (332.01,16490.02) .. (333.68,16491.69) .. controls (335.35,16493.36) and (335.35,16495.02) .. (333.68,16496.69) -- (333.68,16496.69)(330.68,16451.69) .. controls (332.35,16453.36) and (332.35,16455.02) .. (330.68,16456.69) .. controls (329.01,16458.36) and (329.01,16460.02) .. (330.68,16461.69) .. controls (332.35,16463.36) and (332.35,16465.02) .. (330.68,16466.69) .. controls (329.01,16468.36) and (329.01,16470.02) .. (330.68,16471.69) .. controls (332.35,16473.36) and (332.35,16475.02) .. (330.68,16476.69) .. controls (329.01,16478.36) and (329.01,16480.02) .. (330.68,16481.69) .. controls (332.35,16483.36) and (332.35,16485.02) .. (330.68,16486.69) .. controls (329.01,16488.36) and (329.01,16490.02) .. (330.68,16491.69) .. controls (332.35,16493.36) and (332.35,16495.02) .. (330.68,16496.69) -- (330.68,16496.69) ;
\draw    (344,16471) -- (344,16484) ;
\draw [shift={(344,16468)}, rotate = 90] [fill={rgb, 255:red, 0; green, 0; blue, 0 }  ][line width=0.08]  [draw opacity=0] (5.36,-2.57) -- (0,0) -- (5.36,2.57) -- cycle    ;
\path  [shading=_ehsthm2yr,_i1th39cse] (331.68,16491.76) .. controls (326.43,16491.76) and (322.18,16495.67) .. (322.18,16500.5) .. controls (322.18,16505.32) and (326.43,16509.24) .. (331.68,16509.24) .. controls (336.92,16509.24) and (341.17,16505.32) .. (341.17,16500.5) .. controls (341.17,16495.67) and (336.92,16491.76) .. (331.68,16491.76) -- cycle ; 
 \draw   (331.68,16491.76) .. controls (326.43,16491.76) and (322.18,16495.67) .. (322.18,16500.5) .. controls (322.18,16505.32) and (326.43,16509.24) .. (331.68,16509.24) .. controls (336.92,16509.24) and (341.17,16505.32) .. (341.17,16500.5) .. controls (341.17,16495.67) and (336.92,16491.76) .. (331.68,16491.76) -- cycle ; 

\path  [shading=_02783xg5u,_ssoze1dis] (331.5,16491.25) .. controls (336.71,16491.27) and (340.91,16495.33) .. (340.88,16500.31) .. controls (340.85,16505.3) and (336.6,16509.31) .. (331.39,16509.29) .. controls (326.17,16509.26) and (321.97,16505.2) .. (322,16500.22) .. controls (322.03,16495.24) and (326.28,16491.22) .. (331.5,16491.25) -- cycle ; 
 \draw   (331.5,16491.25) .. controls (336.71,16491.27) and (340.91,16495.33) .. (340.88,16500.31) .. controls (340.85,16505.3) and (336.6,16509.31) .. (331.39,16509.29) .. controls (326.17,16509.26) and (321.97,16505.2) .. (322,16500.22) .. controls (322.03,16495.24) and (326.28,16491.22) .. (331.5,16491.25) -- cycle ; 

\draw (382,16436) node [anchor=north west][inner sep=0.75pt]    {$p'$};
\draw (272,16440) node [anchor=north west][inner sep=0.75pt]    {$p$};
\draw (350,16472) node [anchor=north west][inner sep=0.75pt]    {$q$};
\draw (416.67,16468.67) node [anchor=north west][inner sep=0.75pt]    {$=i\mathcal{A}_0(p\to p')\,.$};
\end{tikzpicture}
\]
(The blue gradient of the blob is intended to represent the time dependence of the Vaidya source.) 
The explicit expression reads 
\begin{equation}
    i\mathcal{A}_0(p\to p')= i \int \dd^4 x \, \frac{2GM (t+r)}{r}\bigg[(\partial_t-\partial_r)e^{-ip\cdot x}\bigg]\bigg[(\partial_t-\partial_r)e^{ip'\cdot x}\bigg],
\end{equation}
where the subscript of $\mathcal{A}_0$ indicates a tree-level amplitude. 

The tree amplitude simplifies in the geometric optics limit~\eqref{eq:geolocks}. 
Retaining only the leading term in $\xbar$, \text{i.e.} $p\approx p'$,  the derivatives simply act as 
\begin{equation}\label{der}
   \bigg[(\partial_t-\partial_r)e^{-ip\cdot x}\bigg] \bigg[(\partial_t-\partial_r)e^{ip'\cdot x}\bigg]\approx (E')^2(1+\hat{\vec{r}}\cdot \hat{\vec{p}}')^2\,e^{iq\cdot x}.
\end{equation}
The energy approximation above is familiar in (semi-)classical applications of amplitudes, but it is a critical step in this argument which we will return to in section~\ref{sec:forwardsVbackwards}.
To simplify the notation we drop the prime in the external data $p'$ from here on out. 
Using equation~\eqref{der} we obtain the tree-level leading fragment 
\begin{equation}\label{clfrag}
     i\mathcal{A}^{(0)}_{0}(p-q\to p)= i \int \dd^4 x \, \frac{2GM(t+r)}{r}E^2(1+\hat{\vec{r}}\cdot \hat{\vec{p}})^2\,e^{iq\cdot x }   \equiv i\mathcal{A}^{(0)}_{0}(q)\,.
\end{equation}
Here we are borrowing the ``fragment'' notation\footnote{We only consider 2-point amplitudes in this work, so (unlike reference~\cite{Cristofoli:2021jas}) we do not need any further label to specify the number of external states.} of \cite{Cristofoli:2021jas}: $\mathcal{A}^{(p)}_{L}$ is the $\mathcal{O}(\xbar^p)$ coefficient in the Laurent series of an  $L$-loop amplitude; this is the $p$th fragment of the full amplitude $\mathcal{A}_L$. The notation $i\mathcal{A}^{(0)}_{0}(q)$ underlines the fact that the only  nontrivial dependence on the momentum mismatch $q$ is in the Fourier exponent of the leading fragment.
Furthermore, we also approximate the delta function constraint appearing in equation~\eqref{eq:fullInteraction} in the geometric optics limit, writing
\[\label{eq:deltaApprox}
\hat{\delta}(2 p \cdot q + q^2) \simeq \hat{\delta}(2 p \cdot q) \,.
\]
Again, this is familiar from the KMOC formalism (for example). 
We retain corrections to both of these approximations at the next order in section~\ref{sec:OneLoop}.

Hence, in the geometric optics approximation, we can write, using equation \eqref{clfrag}
\[\label{state1}
\braket{p|{S}_{\text{tree}}-1|\psi}&= 
\int \dd v \,\varphi(v) e^{i p\cdot b(v)}  
\int \dd^4 x \,\hat{\dd}^4 q\, \hat{\delta}(2 p\cdot q)
\\&
\hspace{1cm}
\times
\frac{2iGM(t+r)}{r}E^2(1+\hat{\vec{r}}\cdot \hat{\vec{p}})^2e^{iq\cdot (x-b(v))}.
\]
Our next task is to compute the $x$ and $q$ integrals.
One convenient way to do so is to trivialise them by   Fourier transforming the on-shell delta function inside the integrand. 
For a given  function  of the coordinates $f(x)$  one has
\[\label{ftlambda}
\int  \dd^4 x \,\hat{\dd}^4 q\, \hat{\delta}(2 p\cdot q)f(x) e^{iq\cdot (x-b)}
&=
\int \dd\lambda\,\dd^4 x \,\hat{\dd}^4 q\, f(x)e^{iq\cdot (x-b-2\lambda p)} \\
&=\int \dd\lambda\,f(b+2\lambda p) \,.
\]
Notice that the Fourier transform introduced an affine parameter $\lambda$ along an effective worldline trajectory for the particle $x = b + 2 \lambda p$. 
For the case at hand we have 
\begin{equation}
    f(x)=\frac{M(t+r)}{r}(1+\hat{\vec{r}}\cdot \hat{\vec{p}})^2.
\end{equation}
At this point, as in section~\ref{sec:Review} we choose the particularly simple mass function $M(v) = M(t+r)$ given in equation~\eqref{massv}.
Evaluating the coordinates on the worldline, one finds immediately  
\begin{equation}
  r=2E|\lambda|,\,\,\,\, \,\,\,\,\,t =v+2E\lambda \,.
\end{equation}
Importantly, this  allows us to expose the physical domain of integration which is determined by the black hole's appearance when $t+r>0$. 
The integral is then restricted to be over all  positive $\lambda$-values starting from $-v/(4E)$: 
\begin{equation} \label{eq:intlamtree}
\begin{split}
        \int_{-\infty}^\infty \dd\lambda\,\left[  \frac{M(t+r)}{r}(1+\hat{\vec{r}}\cdot \hat{\vec{p}})^2 \right] _{x=b+2\lambda p}&=\int_{-\infty}^{\infty}\dd\lambda \frac{M\,\Theta(v+2E(\lambda+|\lambda|))4\Theta(\lambda)}{2E|\lambda|} 
       \\& =\frac{2M}{E}\int_{-\frac{v}{4E}}^{\infty} \frac{\dd\lambda}{\lambda} 
        \\&=-\frac{2M}{E}\log(-v/\mu) +\frac{2M}{E}\log (4E\lambda_{\infty}/\mu),
\end{split}
\end{equation}
where $\mu$ is an arbitrary length scale which we introduce on dimensional grounds.
We will discuss the limits on $v$ below.
Note that above we have encountered a large $\lambda$ divergence. 
This is in fact the Coulomb phase we encountered before in equation~\eqref{eq:CoulombPhase}: the logarithmic drag at large $\lambda$ time.
We will see below that the Coulomb phase drops out of observables and so we will neglect it in what follows.

Plugging equations \eqref{ftlambda} and \eqref{eq:intlamtree} inside \eqref{state1}, we are finally able to write down the leading-order result:
\[
\label{treestate}
 \braket{p|{S}_{\text{tree}}-1|\psi}
 =
   \int \dd v \,\varphi(v) e^{i p\cdot b(v)}   \left(-4iGME \log({-v}/\mu) \right).
\]
In the next subsection we will see how the logarithm above exponentiates. 

\subsection{Exponentiation of the leading term}
\label{sec:Exp_Leading}

We have computed the tree contribution to the amplitude $\braket{p|S|\psi}$, but it is very easy to show how this term is resummed into an eikonal-style exponential.  
In order to show this, we begin by considering an $L$-loop comb diagram of the following type
\[   \label{llooop}
\begin{tikzpicture}[x=0.75pt,y=0.75pt,yscale=-1,xscale=1]
  
\tikzset {_glmgh0uvn/.code = {\pgfsetadditionalshadetransform{ \pgftransformshift{\pgfpoint{0 bp } { 0 bp }  }  \pgftransformrotate{0 }  \pgftransformscale{2 }  }}}
\pgfdeclarehorizontalshading{_8zuqzha68}{150bp}{rgb(0bp)=(1,1,1);
rgb(42.606724330357146bp)=(1,1,1);
rgb(45.89285714285714bp)=(0.82,0.89,0.97);
rgb(57.14285714285714bp)=(0.29,0.56,0.89);
rgb(57.25bp)=(0.29,0.56,0.89);
rgb(62.5bp)=(0.29,0.56,0.89);
rgb(100bp)=(0.29,0.56,0.89)}

  
\tikzset {_9cguagxni/.code = {\pgfsetadditionalshadetransform{ \pgftransformshift{\pgfpoint{0 bp } { 0 bp }  }  \pgftransformrotate{0 }  \pgftransformscale{2 }  }}}
\pgfdeclarehorizontalshading{_ppqxu7c63}{150bp}{rgb(0bp)=(1,1,1);
rgb(42.428152901785715bp)=(1,1,1);
rgb(47.67857142857143bp)=(0.82,0.89,0.97);
rgb(52.23214285714286bp)=(0.29,0.56,0.89);
rgb(57.5bp)=(0.29,0.47,0.89);
rgb(62.42815290178571bp)=(0.29,0.47,0.89);
rgb(100bp)=(0.29,0.47,0.89)}

  
\tikzset {_3du76znoo/.code = {\pgfsetadditionalshadetransform{ \pgftransformshift{\pgfpoint{0 bp } { 0 bp }  }  \pgftransformrotate{0 }  \pgftransformscale{2 }  }}}
\pgfdeclarehorizontalshading{_mhwcldo8g}{150bp}{rgb(0bp)=(1,1,1);
rgb(42.606724330357146bp)=(1,1,1);
rgb(45.89285714285714bp)=(0.82,0.89,0.97);
rgb(57.14285714285714bp)=(0.29,0.56,0.89);
rgb(57.25bp)=(0.29,0.56,0.89);
rgb(62.5bp)=(0.29,0.56,0.89);
rgb(100bp)=(0.29,0.56,0.89)}

  
\tikzset {_afv89qnx0/.code = {\pgfsetadditionalshadetransform{ \pgftransformshift{\pgfpoint{0 bp } { 0 bp }  }  \pgftransformrotate{0 }  \pgftransformscale{2 }  }}}
\pgfdeclarehorizontalshading{_rj0fab13l}{150bp}{rgb(0bp)=(1,1,1);
rgb(42.428152901785715bp)=(1,1,1);
rgb(47.67857142857143bp)=(0.82,0.89,0.97);
rgb(52.23214285714286bp)=(0.29,0.56,0.89);
rgb(57.5bp)=(0.29,0.47,0.89);
rgb(62.42815290178571bp)=(0.29,0.47,0.89);
rgb(100bp)=(0.29,0.47,0.89)}

  
\tikzset {_qey65mkkr/.code = {\pgfsetadditionalshadetransform{ \pgftransformshift{\pgfpoint{0 bp } { 0 bp }  }  \pgftransformrotate{0 }  \pgftransformscale{2 }  }}}
\pgfdeclarehorizontalshading{_iec00pnfj}{150bp}{rgb(0bp)=(1,1,1);
rgb(42.606724330357146bp)=(1,1,1);
rgb(45.89285714285714bp)=(0.82,0.89,0.97);
rgb(57.14285714285714bp)=(0.29,0.56,0.89);
rgb(57.25bp)=(0.29,0.56,0.89);
rgb(62.5bp)=(0.29,0.56,0.89);
rgb(100bp)=(0.29,0.56,0.89)}

  
\tikzset {_agax22ioz/.code = {\pgfsetadditionalshadetransform{ \pgftransformshift{\pgfpoint{0 bp } { 0 bp }  }  \pgftransformrotate{0 }  \pgftransformscale{2 }  }}}
\pgfdeclarehorizontalshading{_5y5m57tmm}{150bp}{rgb(0bp)=(1,1,1);
rgb(42.428152901785715bp)=(1,1,1);
rgb(47.67857142857143bp)=(0.82,0.89,0.97);
rgb(52.23214285714286bp)=(0.29,0.56,0.89);
rgb(57.5bp)=(0.29,0.47,0.89);
rgb(62.42815290178571bp)=(0.29,0.47,0.89);
rgb(100bp)=(0.29,0.47,0.89)}

  
\tikzset {_70w6tojmt/.code = {\pgfsetadditionalshadetransform{ \pgftransformshift{\pgfpoint{0 bp } { 0 bp }  }  \pgftransformrotate{0 }  \pgftransformscale{2 }  }}}
\pgfdeclarehorizontalshading{_i9q8704kd}{150bp}{rgb(0bp)=(1,1,1);
rgb(42.606724330357146bp)=(1,1,1);
rgb(45.89285714285714bp)=(0.82,0.89,0.97);
rgb(57.14285714285714bp)=(0.29,0.56,0.89);
rgb(57.25bp)=(0.29,0.56,0.89);
rgb(62.5bp)=(0.29,0.56,0.89);
rgb(100bp)=(0.29,0.56,0.89)}

  
\tikzset {_8sg04z4cg/.code = {\pgfsetadditionalshadetransform{ \pgftransformshift{\pgfpoint{0 bp } { 0 bp }  }  \pgftransformrotate{0 }  \pgftransformscale{2 }  }}}
\pgfdeclarehorizontalshading{_0wpyd7ua6}{150bp}{rgb(0bp)=(1,1,1);
rgb(42.428152901785715bp)=(1,1,1);
rgb(47.67857142857143bp)=(0.82,0.89,0.97);
rgb(52.23214285714286bp)=(0.29,0.56,0.89);
rgb(57.5bp)=(0.29,0.47,0.89);
rgb(62.42815290178571bp)=(0.29,0.47,0.89);
rgb(100bp)=(0.29,0.47,0.89)}
\tikzset{every picture/.style={line width=0.75pt}} 

\draw    (118,16559) -- (164.02,16566.46) ;
\draw [shift={(143.58,16563.15)}, rotate = 189.21] [fill={rgb, 255:red, 0; green, 0; blue, 0 }  ][line width=0.08]  [draw opacity=0] (5.36,-2.57) -- (0,0) -- (5.36,2.57) -- cycle    ;
\draw    (414.82,16565.69) -- (461,16557) ;
\draw [shift={(440.47,16560.87)}, rotate = 169.34] [fill={rgb, 255:red, 0; green, 0; blue, 0 }  ][line width=0.08]  [draw opacity=0] (5.36,-2.57) -- (0,0) -- (5.36,2.57) -- cycle    ;
\draw    (362,16566) -- (414.82,16565.69) ;
\draw [shift={(391.01,16565.83)}, rotate = 179.67] [fill={rgb, 255:red, 0; green, 0; blue, 0 }  ][line width=0.08]  [draw opacity=0] (5.36,-2.57) -- (0,0) -- (5.36,2.57) -- cycle    ;
\draw    (281.63,16566.93) -- (327.61,16566.47) ;
\draw [shift={(307.22,16566.67)}, rotate = 179.43] [fill={rgb, 255:red, 0; green, 0; blue, 0 }  ][line width=0.08]  [draw opacity=0] (5.36,-2.57) -- (0,0) -- (5.36,2.57) -- cycle    ;
\draw    (164.02,16566.46) -- (222.82,16566.69) ;
\draw [shift={(196.02,16566.59)}, rotate = 180.23] [fill={rgb, 255:red, 0; green, 0; blue, 0 }  ][line width=0.08]  [draw opacity=0] (5.36,-2.57) -- (0,0) -- (5.36,2.57) -- cycle    ;
\draw    (222.82,16566.69) -- (281.63,16566.93) ;
\draw [shift={(254.83,16566.82)}, rotate = 180.23] [fill={rgb, 255:red, 0; green, 0; blue, 0 }  ][line width=0.08]  [draw opacity=0] (5.36,-2.57) -- (0,0) -- (5.36,2.57) -- cycle    ;
\draw    (165.28,16566.31) .. controls (166.95,16567.98) and (166.95,16569.64) .. (165.28,16571.31) .. controls (163.61,16572.98) and (163.61,16574.64) .. (165.28,16576.31) .. controls (166.95,16577.98) and (166.95,16579.64) .. (165.28,16581.31) .. controls (163.61,16582.98) and (163.61,16584.64) .. (165.28,16586.31) .. controls (166.95,16587.98) and (166.95,16589.64) .. (165.28,16591.31) .. controls (163.61,16592.98) and (163.61,16594.64) .. (165.28,16596.31) .. controls (166.95,16597.98) and (166.95,16599.64) .. (165.28,16601.31) .. controls (163.61,16602.98) and (163.61,16604.64) .. (165.28,16606.31) .. controls (166.95,16607.98) and (166.95,16609.64) .. (165.28,16611.31) -- (165.28,16611.31)(162.28,16566.31) .. controls (163.95,16567.98) and (163.95,16569.64) .. (162.28,16571.31) .. controls (160.61,16572.98) and (160.61,16574.64) .. (162.28,16576.31) .. controls (163.95,16577.98) and (163.95,16579.64) .. (162.28,16581.31) .. controls (160.61,16582.98) and (160.61,16584.64) .. (162.28,16586.31) .. controls (163.95,16587.98) and (163.95,16589.64) .. (162.28,16591.31) .. controls (160.61,16592.98) and (160.61,16594.64) .. (162.28,16596.31) .. controls (163.95,16597.98) and (163.95,16599.64) .. (162.28,16601.31) .. controls (160.61,16602.98) and (160.61,16604.64) .. (162.28,16606.31) .. controls (163.95,16607.98) and (163.95,16609.64) .. (162.28,16611.31) -- (162.28,16611.31) ;
\draw    (175.6,16585.62) -- (175.6,16598.62) ;
\draw [shift={(175.6,16582.62)}, rotate = 90] [fill={rgb, 255:red, 0; green, 0; blue, 0 }  ][line width=0.08]  [draw opacity=0] (5.36,-2.57) -- (0,0) -- (5.36,2.57) -- cycle    ;
\draw    (224.32,16566.7) .. controls (225.98,16568.37) and (225.97,16570.04) .. (224.3,16571.7) .. controls (222.63,16573.36) and (222.62,16575.03) .. (224.27,16576.7) .. controls (225.92,16578.37) and (225.91,16580.04) .. (224.24,16581.7) .. controls (222.57,16583.36) and (222.56,16585.03) .. (224.21,16586.7) .. controls (225.87,16588.37) and (225.86,16590.04) .. (224.19,16591.7) .. controls (222.52,16593.36) and (222.51,16595.03) .. (224.16,16596.7) .. controls (225.81,16598.37) and (225.8,16600.04) .. (224.13,16601.7) .. controls (222.46,16603.36) and (222.45,16605.03) .. (224.1,16606.7) -- (224.08,16611.32) -- (224.08,16611.32)(221.32,16566.68) .. controls (222.98,16568.35) and (222.97,16570.02) .. (221.3,16571.68) .. controls (219.63,16573.34) and (219.62,16575.01) .. (221.27,16576.68) .. controls (222.92,16578.35) and (222.91,16580.02) .. (221.24,16581.68) .. controls (219.57,16583.34) and (219.56,16585.01) .. (221.21,16586.68) .. controls (222.87,16588.35) and (222.86,16590.02) .. (221.19,16591.68) .. controls (219.52,16593.34) and (219.51,16595.01) .. (221.16,16596.68) .. controls (222.81,16598.35) and (222.8,16600.02) .. (221.13,16601.68) .. controls (219.46,16603.34) and (219.45,16605.01) .. (221.1,16606.68) -- (221.08,16611.3) -- (221.08,16611.3) ;
\draw    (234.4,16585.62) -- (234.4,16598.62) ;
\draw [shift={(234.4,16582.62)}, rotate = 90] [fill={rgb, 255:red, 0; green, 0; blue, 0 }  ][line width=0.08]  [draw opacity=0] (5.36,-2.57) -- (0,0) -- (5.36,2.57) -- cycle    ;
\draw    (283.13,16566.94) .. controls (284.78,16568.61) and (284.77,16570.28) .. (283.1,16571.94) .. controls (281.43,16573.6) and (281.42,16575.27) .. (283.07,16576.94) .. controls (284.73,16578.61) and (284.72,16580.28) .. (283.05,16581.94) .. controls (281.38,16583.6) and (281.37,16585.27) .. (283.02,16586.94) .. controls (284.67,16588.61) and (284.66,16590.28) .. (282.99,16591.94) .. controls (281.32,16593.6) and (281.31,16595.27) .. (282.96,16596.94) .. controls (284.61,16598.61) and (284.6,16600.28) .. (282.93,16601.94) .. controls (281.26,16603.6) and (281.25,16605.27) .. (282.9,16606.94) .. controls (284.56,16608.61) and (284.55,16610.28) .. (282.88,16611.94) -- (282.88,16612.12) -- (282.88,16612.12)(280.13,16566.92) .. controls (281.78,16568.59) and (281.77,16570.26) .. (280.1,16571.92) .. controls (278.43,16573.58) and (278.42,16575.25) .. (280.07,16576.92) .. controls (281.73,16578.59) and (281.72,16580.26) .. (280.05,16581.92) .. controls (278.38,16583.58) and (278.37,16585.25) .. (280.02,16586.92) .. controls (281.67,16588.59) and (281.66,16590.26) .. (279.99,16591.92) .. controls (278.32,16593.58) and (278.31,16595.25) .. (279.96,16596.92) .. controls (281.61,16598.59) and (281.6,16600.26) .. (279.93,16601.92) .. controls (278.26,16603.58) and (278.25,16605.25) .. (279.91,16606.92) .. controls (281.56,16608.59) and (281.55,16610.26) .. (279.88,16611.92) -- (279.88,16612.1) -- (279.88,16612.1) ;
\draw    (293.2,16586.42) -- (293.2,16599.42) ;
\draw [shift={(293.2,16583.42)}, rotate = 90] [fill={rgb, 255:red, 0; green, 0; blue, 0 }  ][line width=0.08]  [draw opacity=0] (5.36,-2.57) -- (0,0) -- (5.36,2.57) -- cycle    ;
\draw    (416.32,16565.69) .. controls (417.99,16567.35) and (418,16569.02) .. (416.34,16570.69) .. controls (414.68,16572.36) and (414.69,16574.03) .. (416.36,16575.69) .. controls (418.03,16577.36) and (418.03,16579.02) .. (416.37,16580.69) .. controls (414.71,16582.36) and (414.72,16584.03) .. (416.39,16585.69) .. controls (418.06,16587.35) and (418.07,16589.02) .. (416.41,16590.69) .. controls (414.75,16592.36) and (414.75,16594.02) .. (416.42,16595.69) .. controls (418.09,16597.35) and (418.1,16599.02) .. (416.44,16600.69) .. controls (414.78,16602.36) and (414.79,16604.03) .. (416.46,16605.69) .. controls (418.13,16607.35) and (418.14,16609.02) .. (416.48,16610.69) -- (416.48,16610.91) -- (416.48,16610.91)(413.32,16565.7) .. controls (414.99,16567.36) and (415,16569.03) .. (413.34,16570.7) .. controls (411.68,16572.37) and (411.69,16574.04) .. (413.36,16575.7) .. controls (415.03,16577.37) and (415.03,16579.03) .. (413.37,16580.7) .. controls (411.71,16582.37) and (411.72,16584.04) .. (413.39,16585.7) .. controls (415.06,16587.36) and (415.07,16589.03) .. (413.41,16590.7) .. controls (411.75,16592.37) and (411.75,16594.03) .. (413.42,16595.7) .. controls (415.09,16597.36) and (415.1,16599.03) .. (413.44,16600.7) .. controls (411.78,16602.37) and (411.79,16604.04) .. (413.46,16605.7) .. controls (415.13,16607.36) and (415.14,16609.03) .. (413.48,16610.7) -- (413.48,16610.92) -- (413.48,16610.92) ;
\draw    (426.8,16585.22) -- (426.8,16598.22) ;
\draw [shift={(426.8,16582.22)}, rotate = 90] [fill={rgb, 255:red, 0; green, 0; blue, 0 }  ][line width=0.08]  [draw opacity=0] (5.36,-2.57) -- (0,0) -- (5.36,2.57) -- cycle    ;
\path  [shading=_8zuqzha68,_glmgh0uvn] (163.58,16605.71) .. controls (158.93,16605.71) and (155.16,16609.23) .. (155.16,16613.58) .. controls (155.16,16617.93) and (158.93,16621.46) .. (163.58,16621.46) .. controls (168.23,16621.46) and (172,16617.93) .. (172,16613.58) .. controls (172,16609.23) and (168.23,16605.71) .. (163.58,16605.71) -- cycle ; 
 \draw   (163.58,16605.71) .. controls (158.93,16605.71) and (155.16,16609.23) .. (155.16,16613.58) .. controls (155.16,16617.93) and (158.93,16621.46) .. (163.58,16621.46) .. controls (168.23,16621.46) and (172,16617.93) .. (172,16613.58) .. controls (172,16609.23) and (168.23,16605.71) .. (163.58,16605.71) -- cycle ; 

\path  [shading=_ppqxu7c63,_9cguagxni] (163.42,16605.25) .. controls (168.04,16605.27) and (171.77,16608.93) .. (171.74,16613.42) .. controls (171.72,16617.9) and (167.95,16621.52) .. (163.32,16621.5) .. controls (158.7,16621.48) and (154.97,16617.82) .. (155,16613.33) .. controls (155.03,16608.84) and (158.8,16605.22) .. (163.42,16605.25) -- cycle ; 
 \draw   (163.42,16605.25) .. controls (168.04,16605.27) and (171.77,16608.93) .. (171.74,16613.42) .. controls (171.72,16617.9) and (167.95,16621.52) .. (163.32,16621.5) .. controls (158.7,16621.48) and (154.97,16617.82) .. (155,16613.33) .. controls (155.03,16608.84) and (158.8,16605.22) .. (163.42,16605.25) -- cycle ; 

\path  [shading=_mhwcldo8g,_3du76znoo] (222.58,16605.71) .. controls (217.93,16605.71) and (214.16,16609.23) .. (214.16,16613.58) .. controls (214.16,16617.93) and (217.93,16621.46) .. (222.58,16621.46) .. controls (227.23,16621.46) and (231,16617.93) .. (231,16613.58) .. controls (231,16609.23) and (227.23,16605.71) .. (222.58,16605.71) -- cycle ; 
 \draw   (222.58,16605.71) .. controls (217.93,16605.71) and (214.16,16609.23) .. (214.16,16613.58) .. controls (214.16,16617.93) and (217.93,16621.46) .. (222.58,16621.46) .. controls (227.23,16621.46) and (231,16617.93) .. (231,16613.58) .. controls (231,16609.23) and (227.23,16605.71) .. (222.58,16605.71) -- cycle ; 

\path  [shading=_rj0fab13l,_afv89qnx0] (222.42,16605.25) .. controls (227.04,16605.27) and (230.77,16608.93) .. (230.74,16613.42) .. controls (230.72,16617.9) and (226.95,16621.52) .. (222.32,16621.5) .. controls (217.7,16621.48) and (213.97,16617.82) .. (214,16613.33) .. controls (214.03,16608.84) and (217.8,16605.22) .. (222.42,16605.25) -- cycle ; 
 \draw   (222.42,16605.25) .. controls (227.04,16605.27) and (230.77,16608.93) .. (230.74,16613.42) .. controls (230.72,16617.9) and (226.95,16621.52) .. (222.32,16621.5) .. controls (217.7,16621.48) and (213.97,16617.82) .. (214,16613.33) .. controls (214.03,16608.84) and (217.8,16605.22) .. (222.42,16605.25) -- cycle ; 

\path  [shading=_iec00pnfj,_qey65mkkr] (281.58,16605.71) .. controls (276.93,16605.71) and (273.16,16609.23) .. (273.16,16613.58) .. controls (273.16,16617.93) and (276.93,16621.46) .. (281.58,16621.46) .. controls (286.23,16621.46) and (290,16617.93) .. (290,16613.58) .. controls (290,16609.23) and (286.23,16605.71) .. (281.58,16605.71) -- cycle ; 
 \draw   (281.58,16605.71) .. controls (276.93,16605.71) and (273.16,16609.23) .. (273.16,16613.58) .. controls (273.16,16617.93) and (276.93,16621.46) .. (281.58,16621.46) .. controls (286.23,16621.46) and (290,16617.93) .. (290,16613.58) .. controls (290,16609.23) and (286.23,16605.71) .. (281.58,16605.71) -- cycle ; 

\path  [shading=_5y5m57tmm,_agax22ioz] (281.42,16605.25) .. controls (286.04,16605.27) and (289.77,16608.93) .. (289.74,16613.42) .. controls (289.72,16617.9) and (285.95,16621.52) .. (281.32,16621.5) .. controls (276.7,16621.48) and (272.97,16617.82) .. (273,16613.33) .. controls (273.03,16608.84) and (276.8,16605.22) .. (281.42,16605.25) -- cycle ; 
 \draw   (281.42,16605.25) .. controls (286.04,16605.27) and (289.77,16608.93) .. (289.74,16613.42) .. controls (289.72,16617.9) and (285.95,16621.52) .. (281.32,16621.5) .. controls (276.7,16621.48) and (272.97,16617.82) .. (273,16613.33) .. controls (273.03,16608.84) and (276.8,16605.22) .. (281.42,16605.25) -- cycle ; 

\path  [shading=_i9q8704kd,_70w6tojmt] (414.58,16606.71) .. controls (409.93,16606.71) and (406.16,16610.23) .. (406.16,16614.58) .. controls (406.16,16618.93) and (409.93,16622.46) .. (414.58,16622.46) .. controls (419.23,16622.46) and (423,16618.93) .. (423,16614.58) .. controls (423,16610.23) and (419.23,16606.71) .. (414.58,16606.71) -- cycle ; 
 \draw   (414.58,16606.71) .. controls (409.93,16606.71) and (406.16,16610.23) .. (406.16,16614.58) .. controls (406.16,16618.93) and (409.93,16622.46) .. (414.58,16622.46) .. controls (419.23,16622.46) and (423,16618.93) .. (423,16614.58) .. controls (423,16610.23) and (419.23,16606.71) .. (414.58,16606.71) -- cycle ; 

\path  [shading=_0wpyd7ua6,_8sg04z4cg] (414.42,16606.25) .. controls (419.04,16606.27) and (422.77,16609.93) .. (422.74,16614.42) .. controls (422.72,16618.9) and (418.95,16622.52) .. (414.32,16622.5) .. controls (409.7,16622.48) and (405.97,16618.82) .. (406,16614.33) .. controls (406.03,16609.84) and (409.8,16606.22) .. (414.42,16606.25) -- cycle ; 
 \draw   (414.42,16606.25) .. controls (419.04,16606.27) and (422.77,16609.93) .. (422.74,16614.42) .. controls (422.72,16618.9) and (418.95,16622.52) .. (414.32,16622.5) .. controls (409.7,16622.48) and (405.97,16618.82) .. (406,16614.33) .. controls (406.03,16609.84) and (409.8,16606.22) .. (414.42,16606.25) -- cycle ; 

\draw (499,16583) node [anchor=north west][inner sep=0.75pt]    {$=i \mathcal{A}_L(p\to p').$};
\draw (103,16554) node [anchor=north west][inner sep=0.75pt]    {$p$};
\draw (466,16549) node [anchor=north west][inner sep=0.75pt]    {$p'$};
\draw (334,16562) node [anchor=north west][inner sep=0.75pt]    {$\cdots $};
\draw (180.6,16584.62) node [anchor=north west][inner sep=0.75pt]    {$\ell _{1}$};
\draw (239.4,16584.62) node [anchor=north west][inner sep=0.75pt]    {$\ell _{2}$};
\draw (298.2,16585.42) node [anchor=north west][inner sep=0.75pt]    {$\ell _{3}$};
\draw (430.8,16586.22) node [anchor=north west][inner sep=0.75pt]    {$\ell _{L+1}$};

\end{tikzpicture}
\] 
As the notation suggests, the diagram above can be viewed as an $L$-loop graph; pictorially this follows from joining the sources together as a line.
Indeed this is the only $L$-loop diagram in the theory of a scalar coupled to a Kerr-Schild background\footnote{Diagrams with crossed graviton legs are equal to their uncrossed cousins in the probe limit.}. 

The leading order fragment of \eqref{llooop} in the geometric optics approximation corresponds to linearising all internal propagators and in neglecting powers of $\ell_i, q$ in  numerators (a step which we will relax in section \ref{sec:OneLoop}). It is then easy to verify that the leading $L$-loop fragment can be written as,  defining $\ell_{i\cdots j}=\sum_{k=i}^j \ell_k$
\[ \label{sclassL}
    i\mathcal{A}_L^{(-L)}(p &-q \to p)=i^L\int \hat{\dd}^4\ell_1\cdots\hat{\dd}^4\ell_{L+1}\,\hat{\delta}^4(\ell_{12\cdots L+1}-q)
    \\&
    \hspace{4cm}
    \times
    \frac{i \mathcal{A}_0^{(0)}(\ell_1)}{2p\cdot\ell_1+i\epsilon} \frac{i \mathcal{A}_0^{(0)}(\ell_2)}{2p\cdot\ell_{12}+i\epsilon}\cdots \frac{i \mathcal{A}_0^{(0)}(\ell_{L+1})}{2p\cdot\ell_{12\cdots L}+i\epsilon}.
\]
where the fragments $i\mathcal{A}_0^{(0)}(\ell_i)$ were defined earlier in eq. \eqref{clfrag}.
The structure of the product in equation~\eqref{sclassL} is particularly simple in the leading geometric optics approximation: this is because each internal $i\mathcal{A}_0^{(0)}(\ell_i)$ is simply given by a tree-level expression. These insertions are all equal (up to a different integration label $x_i$) and depend on the loop variable only through Fourier phases. 
That is
\begin{equation}\label{prod}
\begin{split}
     i \mathcal{A}_0^{(0)} (\ell_1) \cdots i \mathcal{A}_0^{(0)}(\ell_{L+1}) =
     (2iG E^2)^{L+1}\prod_{i=1}^{L+1} \int \dd^4 x_i \, \frac{M(t_i+r_i)}{r_i} (1+\hat{\vec{r}}_i\cdot \hat{\vec{p}})^2\,e^{i\ell_i\cdot x_i }.
\end{split}
\end{equation}
Let us move onto the propagators next, for these we will need the eikonal identity (see for example~\cite{Akhoury:2013yua})
\[\label{eikid}
\sum_\sigma \frac{i^L\,\hat{\delta}(p \cdot \ell_{1\cdots L+1})}{(p \cdot \ell_{\sigma(1)} +i \epsilon) \cdots ( p \cdot \ell_{\sigma(1) \cdots \sigma(L)}+ i\epsilon)} = \prod_{i=1}^{L+1} \hat{\delta}(p \cdot \ell_i),
\]
where the sum is over permutations of the loop variables.
To make use of this formula we integrate \eqref{sclassL} over $q$ to work in impact parameter  space:
\[
 i\tilde{\mathcal{A}}_L^{(-L)}(b)= \int \hat{\dd}^4 q\, \hat{\delta}(2 p\cdot q) \, i\mathcal{A}^{(-L)}_{L}(p-q\to p) \, e^{-iq\cdot b(v)},
\] 
In fact, the Fourier transform shows that, after integrating out $q$ with ${\hat{\delta}^4(\ell_{1\cdots L+1}-q)}$,  we are left with a permutationally-invariant expression,\footnote{To notice this one has to relabel the dummy position variables $x_i$.} so we can apply equation~\eqref{eikid} at the only cost of a combinatorial factor. Then, the loop function that we are dealing with becomes 
\[ 
\int \hat{\dd}^4\ell_1\, e^{-i\ell_1\cdot(b-x_1)}&\cdots\hat{\dd}^4\ell_{L+1}\,e^{-i\ell_{L+1}\cdot (b-x_{L+1})}\hat{\delta}(2p\cdot \ell_{1\cdots L+1}) 
    \frac{i}{2p\cdot\ell_1+i\epsilon}  \cdots \frac{i}{2p\cdot\ell_{12\cdots L}+i\epsilon}
    \\&\hspace{4cm}
    =\frac{1}{(L+1)!}
\prod_{i=1}^{L+1}\int \hat{\dd}^4\ell_i\, e^{-i\ell_i\cdot (b-x_i)} \hat{\delta}(2p \cdot \ell_i),
\]
which is indeed a pure product in position space.
This fact combined with \eqref{prod} yields the desired exponentiation pattern for the leading amplitude \eqref{sclassL}, apparent after summing over loops 
\begin{align}
\label{eq:ResumedAmp}
 \braket{p|S|\psi}
    &= \int \dd v \,\varphi(v) e^{i p\cdot b(v)} \times\left[1+\sum_{L=0}^\infty\int \hat{\dd}^4 q\, \hat{\delta}(2 p\cdot q)i\mathcal{A}^{(-L)}_{L}(p-q\to p)e^{-iq\cdot b(v)}\right]\nn\\
    &=
     \int \dd v \,\varphi(v) e^{i p\cdot b(v)} \nn\\&\hspace{2.5cm}\times\left[ 1+\sum_{L=0}^\infty\frac{1}{(L+1)!}
    \left(\int \hat{\dd}^4 \ell\, \hat{\delta}(2 p\cdot \ell)i\mathcal{A}^{(0)}_{0}(\ell)e^{-i\ell\cdot b(v)}\right)^{L+1}\right] \nn\\
    &=
    \int \dd v \,\varphi(v) e^{i p\cdot b(v)} e^{-4iGME\log(-v/\mu)}.
\end{align}
Note that in the last line we used the earlier steps \eqref{ftlambda} and \eqref{eq:intlamtree} to explicitly integrate the eikonal phase. This is the main result of the section: the exponential resummation of the leading $L$-loop diagrams.  

Before we move on, let us briefly comment on subleading corrections to formula \eqref{eq:ResumedAmp}. Indeed, in section~\ref{sec:Review} we saw that the Hawking amplitude begins to display  corrections of the order of the horizon radius $\mathcal{O}\left((GM)^2\right)$, which arise from the expansion of the logarithm in equation \eqref{eq:Ahawkfull}. 
Motivated by this, in section~\ref{sec:OneLoop} we will show that our diagrammatic approach can indeed recover this correction correctly at one loop. 
However we are already in a position to compute the Hawking spectrum so let us deal with that important topic first.

\subsection{Hawking's amplitude and the thermal spectrum}\label{spectrum}

Building on the discussion from the previous section, we now define the Hawking amplitude simply as 
\[
\label{eq:HawkingAmp_definition}
    \mathcal{A} (p) 
    \equiv 
    \langle p|{S}|\psi\rangle = 
    \int \! \dd v \, e^{iEv}
    \varphi(v)e^{-4iGM E\log(-v/\mu)} \,.
\]
A first observation is that the branch cut in the logarithm forces us to restrict the region of integration.
The integral in equation~\eqref{eq:HawkingAmp_definition} only makes sense for $v < 0$; for  {positive} values of $v$ our computation evidently breaks down.
This is the first appearance of the horizon in our approach\footnote{We will see that subdominant loop corrections are relevant to the location of the horizon below.}.
We therefore refine our definition to
\[\label{eq:aHgeneraltree}
\mathcal{A} (p) 
    \equiv 
    \braket{p|S|\psi} = 
    \int_{-\infty}^0 \dd v \, e^{iEv}
    \varphi(v)e^{-4iGM E\log(-v/\mu)} \,.
\]
Physically, the contribution of the portion of the wavepacket with $v > 0$ simply enters the black hole and (as far as our current treatment goes) is bound there. 
In other words the $S$ matrix acting on this $v>0$ portion does not yield a state at $\scri^+$.
Since a detector at $\scri^+$ is insensitive to this portion of the incoming wavepacket, our prescription of cutting off the integral is equivalent to tracing out these modes.

At this stage it is useful to pick a definite incoming state $\ket{\psi}$. 
Physically, a wavepacket localising the beam is a very sensible choice, and would again make contact with the KMOC setup.
A wavepacket localised in time would also help us track the time dependence of the scattering (and later of the related pair-production process).
However pragmatically the simplest option is to take $\ket\psi$ to be monochromatic and spherically symmetric, which we can accomplish by setting
\[
\label{eq:InitialStatePsi}
|\psi\rangle 
=
 \int  \frac{\dd \Omega_{{{p}}} }{4 \pi}\ket{E_0, E_0 \hat{\vec{p}}} \,.
\]
As we will soon see, this choice allows us to perform the relevant $v$ integral explicitly.
This state can be understood as a partial wave state with total angular momentum quantum number $\ell = 0$, and is normalised according to the conventions for partial wave states outlined briefly in appendix~\ref{app:sphericalNorms};
see reference~\cite{Aoude:2023fdm} for a far more thorough discussion of this class of states in the modern context of scattering amplitudes.
Our choice of $\ket{\psi}$ then requires
\[
\label{eq:InitialWaveFunction}
    \varphi(v) = \frac{2 \pi}{E_0}\, e^{-iE_0v}  \,.
\]
Recall that we also chose a spherically symmetric energy eigenfunction in equation~\eqref{eq:incomingWavefunction}, so we can compare the amplitudes in equations~\eqref{eq:AHawking} and~\eqref{eq:aHgeneraltree}.
 
As for the bra $\bra{p}$ in equation~\eqref{eq:aHgeneraltree}, we should also ensure agreement with the discussion of section~\ref{sec:Review} --- there, both initial and final wavefunctions were taken to be spherically symmetric. 
However, both the $S$ matrix and the ket $\ket{\psi}$ in equation~\eqref{eq:aHgeneraltree} are spherically symmetric, and hence $\mathcal{A}$ (manifestly) does not depend on the angles of ${\bm p}$.
The angular averaging of this bra is therefore trivial, and the conventions of reference~\ref{app:sphericalNorms} guarantee that no additional normalisation factor arises.
We write the angular-averaged amplitude as
\[\label{eq:aHspherical}
\mathcal{A} (E) 
= \frac{2\pi}{E_0} \int_{-\infty}^0 \dd v \, e^{i(E-E_0) v}
    e^{-4iGM E\log(-v/\mu)} \,.
\]
We can now directly compare equations~\eqref{eq:AHawking} and~\eqref{eq:aHspherical}.
The two equations agree at leading order, up to normalisation which we discuss in section~\ref{sec:forwardsVbackwards}.

We now recognise that the $v$ integral in~\eqref{eq:aHspherical} is a complete gamma function, so that\footnote{Strictly speaking one has to ensure convergence of the $v$ integral via  $(E-E_0)\to (E-E_0)-i\epsilon$. Since this will not be important for our purposes we omit the $i\epsilon$. We have also dropped an irrelevant constant phase.}
\[ 
\label{amplih}
\mathcal{A} (E) = \frac{2\pi}{E_0}(i(E-E_0))^{-1+4iGME}\,\Gamma(1-4iGM E).
\] 
As we will see in section \ref{sec:Bogoliubov}, the differential number spectrum, defined as the integrand of $\braket{\vac |S^\dagger a^\dagger a S | \vac}$, is given in terms of this object by
\[\label{eq:dnDistn}
\dd n &= \frac{E}{4\pi^2} \dd E \, \frac{E_0}{4\pi^2}  \dd E_0 \frac{E_0^2}{E^2} |\mathcal{A}(E)|^2 \\
&= \dd E \dd E_0 \frac{1}{4 \pi^2 E E_0}e^{4\pi GME}|\Gamma(1-4 iGME)|^2,
\]
taking $E_0 \gg E$. 
It is at this point that the infrared divergent Coulomb phase we encountered in equation~\eqref{eq:intlamtree} cancels.
The infrared divergence is a $v$ independent phase, and so it drops out of $|\mathcal{A}(E)|^2$.
Similarly the arbitrary scale $\mu$ cancels from the observable $\dd n$ for the same reason\footnote{We thank Fei Teng for clarifying comments on these points.}.

The physics of the number distribution becomes much clearer after use of Euler's reflection formula\footnote{This is $\Gamma(z)\Gamma(1-z)={\pi}/{\sin\pi z}$.}, leading to
\[\label{eq:thermalDist}
\dd n = \dd E \, \dd E_0 \frac{2 G M}{\pi E_0} \frac{e^{8 \pi GM E}}{e^{8\pi GM E}-1} \,.
\]
We have obtained a thermal distribution with temperature
\begin{equation}
    T_{H} =\frac{1}{ 8\pi GM}.
\end{equation}
Notice that the total number of emitted particles is divergent.
As Hawking demonstrated~\cite{Hawking:1975vcx} in detail, this divergence is due to a steady emission of particles over an infinite time; it is therefore an artifact of our approximations.

Of course, it comes as a surprise that squaring a $1\to 1$ amplitude leads to a thermal distribution, even with knowledge of Hawking's work.
We will clarify this phenomenon in the next section by explaining how Bogoliubov coefficients relate to the diagrams we resummed to compute this amplitude, and we will also better understand the factor $e^{8 \pi GM E}$ in the numerator of the thermal distribution~\eqref{eq:thermalDist}.
For now, let us simply note that the combination of exponentials in the distribution~\eqref{eq:thermalDist} can be written in the suggestive form 
\[
\frac{e^{8 \pi GM E}}{e^{8\pi GM E}-1} =1+\frac{1}{e^{8\pi GM E}-1}.
\]
Evidently the last term above is the standard Bose-Einstein distribution; in that sense what remains for us to understand is the origin of the contribution of unity on the right-hand-side above.

\section{Bogoliubov coefficients and amplitudes}
\label{sec:Bogoliubov}

Many discussions of Hawking radiation, following Hawking's original treatment~\cite{Hawking:1975vcx}, use the language of Bogoliubov transformations.
As we will now see there is a very close relationship between these transformations and generalisations of amplitudes which have recently received attention in the literature.

\subsection{The transformations from mode functions}

Let us set the scene with a general review of Bogoliubov transformations in an interacting scalar theory following reference~\cite{Wald:1975kc}.
We consider a situation in which a Gaussian scalar interacts with a background which vanishes in the far past and the far future.
For a gravitationally coupled scalar, the action of the theory is
\[\label{eq:gaussianAction}
S_\textrm{action} = \frac 12 \int \dd^4 x \sqrt{-g} \Big( g^{\mu\nu} \, \partial_\mu \phi \, \partial_\nu \phi - m^2 \phi^2 \Big) \,,
\]
though much of our discussion applies equally to the flat space theory
\[
S_\textrm{action} = \frac 12 \int \dd^4 x\Big(  (\partial \phi)^2 - m^2 \phi^2 - J(x) \phi^2\Big)\,,
\]
where $J(x)$ is an external source. The mass $m$ will play no particular role in this section, so we have included it for completeness.

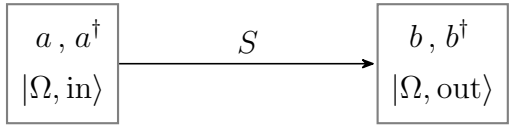
\begin{figure}[t]
\centering
\begin{tikzpicture}[auto]
\node (in) at (0,0) [rectangle,draw=black!50,thick,inner sep=5pt] {$\begin{aligned} a \,, &\; a^\dagger \\ |\vac, \, &\textrm{in} \rangle  \end{aligned}$};
\node (out) at (5,0) [rectangle,draw=black!50,thick,inner sep=5pt] {$\begin{aligned} b \,, & \; b^\dagger \\ |\vac, \, &\textrm{out} \rangle \end{aligned}$};
\draw [->,shorten >=1pt,>={Stealth[round]},semithick] (in.east) to node {$S$} (out.west);
\end{tikzpicture}
\caption{The $S$ matrix maps the in Fock space to the out Fock space.}\label{fig:aSb}
\end{figure}

We choose conventions to match the scattering amplitudes literature as closely as possible, outlined in figure~\ref{fig:aSb}.
As usual, we treat the incoming and outgoing Fock spaces as free particle spaces, and so we expand our quantum scalar field as
\[\label{eq:asymPhiPast}
\phi(x)= \phi_\textrm{in} = \int \dd \Phi(k) \left( e^{-i k \cdot x} a(k) + e^{i k \cdot x} a^\dagger(k) \right) 
\]
in the far past, as $t \rightarrow -\infty$. On the other hand, the field is
\[\label{eq:asymPhiFuture}
\phi(x)= \phi_\textrm{out} = \int \dd \Phi(k) \left( e^{-i k \cdot x} b(k) + e^{i k \cdot x} b^\dagger(k) \right) 
\]
as $t \rightarrow +\infty$.
The incoming ladder operators satisfy $a(k) \ket{\vac, \textrm{in}} = 0$ while the outgoing vacuum is annihilated by the $b(k)$ operators.
Of course these are ``vacua'' in the presence of a non-trivial background.

Since the $S$ matrix relates the incoming and outgoing Fock spaces, we know that
\[
\phi_\textrm{out} = S^\dagger \phi_\textrm{in} S \,.
\]
Orthogonality of the mode expansion for $\phi(x)$ therefore relates the ladder operators:
\[\label{eq:allTimeEvolution}
b(k) = S^\dagger a(k) S \,.
\]
It is important to realise that this does not imply any relationship between $S \ket{\vac, \textrm{in}}$ and $\ket{\vac, \textrm{out}}$.
When possible we will prefer the incoming ladder operators, and therefore we write $\ket{\vac} = \ket{\vac, \textrm{in}}$.

The Gaussian structure of the theory~\eqref{eq:gaussianAction} allows us to go much further.
The equation of motion for the scalar field is linear, and therefore we can (in principle) solve the equation of motion in terms of two linearly independent sets of mode solutions.
We choose Past modes $P(x, k)$ so that as $x^0 \rightarrow -\infty$
\[
P(x, k) \simeq e^{-i k \cdot x} \,,
\]
and a separate Future basis $F(x, k)$ so that as $x^0 \rightarrow +\infty$
\[\label{eq:futureBC}
F(x, k) \simeq e^{-i k \cdot x} \,.
\]
This means we have two equally valid expressions for our scalar field at all times:
\begin{subequations}\label{eq:twoPhis}
\begin{align}
\phi(x) &= \int \dd \Phi(k) \left( P(x, k) a(k) + \bar{P}(x, k) a^\dagger(k) \right) \label{eq:phiPast} \\
&= \int \dd \Phi(k) \left( F(x, k) b(k) + \bar{F}(x, k) b^\dagger(k) \right) \label{eq:phiFuture}\,.
\end{align}
\end{subequations}
Notice that these expressions recover the asymptotic expansions~\eqref{eq:asymPhiPast} and~\eqref{eq:asymPhiFuture} in the past and future, respectively.

It is clear from equation~\eqref{eq:twoPhis} that the two sets of ladder operators are linear combinations of one another.
To see how this works, it is very useful to introduce the Klein-Gordon inner product, acting on the mode functions.
We define
\[
( f(x) | g(x) ) = \int \dd^3 x \, \big( \bar f(x) \, i \partial_t g(x) - g(x) \, i \partial_t \bar f(x) \big) \,.
\]
The integral is to be taken over a three-dimensional Cauchy surface at some time, and it is straightforward to show that the integral is independent of the choice of time because $f(x)$ and $g(x)$ satisfy the Klein-Gordon equation.
Evaluating the inner product for fixed large time we see that
\[ \label{eq:KGortho}
(P(x, k) | P(x, k')) &= \delta_\Phi(k - k') = (F(x, k) | F(x, k')) \,, \\
(P(x, k) | \bar P(x, k')) &= 0  = (F(x, k) | \bar F(x,k') )\,.
\]
The inner product also has the useful properties
\[\label{eq:normProperties}
(f | g) &= (g | f)^* \,,\\
(f | \bar g) &= - (g | \bar f) \,,
\]
and separates the space of solutions of the Klein-Gordon equation into positive- and negative-frequency parts. 
That is, if we write a real generic solution as
\[
&f(x) = \int \dd\Phi(p) \left( e^{-i p \cdot x} {\tilde f}(p) + e^{i p \cdot x} {\tilde  f}^*(p) \right) \,, 
\]
then the inner product with the free mode $e^{-i k \cdot x}$ for $k^0 > 0$ is
\[\label{eq:positiveFreqPart}
\left(e^{-i k \cdot x} | f(x) \right) = {\tilde  f}(k) \,.
\]
The orthogonality properties~\eqref{eq:KGortho} allow us to project the quantum fields on to creation and annihilation operators in a simple way.
For example, using equation~\eqref{eq:phiFuture} we immediately see that
\[
(F(x, k) | \phi(x)  ) = b(k) \,.
\]
Alternatively we can write the quantum field in the inner product above using equation~\eqref{eq:phiPast}, leading to the anticipated relation between the ladder operators:
\[\label{eq:bGivenBya}
b(k) = S^\dagger a(k) S = A(k, p) a(p) + B(k, p) a^\dagger(p) \,.
\]
In this equation we are using a condensed notation, where repeated momenta ($p$ in this case) are to ``summed over'', that is integrated against the on-shell measure. 
In more detail,
\[\label{relar}
b(k) = \int \dd\Phi(p) \big( A(k, p) a(p) + B(k, p) a^\dagger(p) \big) \,.
\]
We have also defined the Bogoliubov coefficients, which play a starring role in this story, as
\[\label{eq:BcoeffsDef}
A(k, p) = (F(x, k) | P(x, p)) \,, \quad B(k, p) = (F(x, k) | \bar P(x, p) ) \,.
\]
These coefficients have a natural matrix structure which motivates our suppression of the integral sign: 
it is convenient to use matrix notation as much as possible.

With knowledge of the Bogoliubov coefficients, we can solve for the dynamics of the theory. 
As an example, we construct the future of the in-vacuum $S \ket{\vac}$. 
First, project $\phi(x)$ onto the $P$ modes and use equation~\eqref{eq:normProperties} to show
\[\label{eq:aGivenByb}
a(p) = A^\dagger(p, k) b(k) - B^T(p, k) b^\dagger(k) \,,
\]
where we have defined
\[
A^\dagger(p, k) = A^*(k, p) \, ,\quad B^T(p, k) = B(k, p) \,.
\]
(These are just the matrix adjoint of $A$ and transpose of $B$ respectively.)
The simple fact that $a(p) \ket{\vac} = 0$ now leads via equations~\eqref{eq:allTimeEvolution} and~\eqref{eq:aGivenByb} to the non-trivial relation
\[\label{eq:vacuumRelation}
a(p) S \ket{\vac} =  \xi (p, k) a^\dagger(k) S \ket{\vac} \,,
\]
where the matrix $\xi$ is defined as
\[
\xi(p, k) = (A^\dagger)^{-1}(p, q) B(q, k) \,.
\]
(Note that the repeated momentum $q$ above is to be integrated over, and the inverse indicated here is a matrix inverse, regarding the momenta as matrix indices.)
It is easy to check that the solution of equation~\eqref{eq:vacuumRelation} is
\[\label{eq:squeezedVacuum}
S \ket{\vac} = \mathcal{N}_\vac \exp \left[ \frac 12 a^\dagger(p) \xi (p, k) a^\dagger(k) \right] \ket{\vac} \,,
\]
where $\mathcal{N}_\vac$ is a normalisation constant\footnote{$\mathcal{N}_\vac$ can be computed from the Baker–Campbell–Hausdorff formula, for a diagrammatic interpretation in terms of vacuum transitions see reference~\cite{Copinger:2024pai}.}. 

The expression~\eqref{eq:squeezedVacuum} contains a lot of interesting physics. 
First, we remark that $ \xi$ must be a symmetric matrix of its arguments~\cite{Wald:1975kc} in view of the  Bose symmetry of $a^\dagger(k)$. 
We see that the past vacuum evolves to a non-trivial state containing correlated particle pairs which are produced by the time-dependent background.
This is a ``squeezing'' process, in the sense of quantum optics \cite{Gerry_Knight_2004}. 
Squeezed states are also of interest in various areas of particle physics and cosmology \cite{PhysRev.183.1057,PhysRevD.3.346,SCHUMAKER1986317,Grishchuk:1990bj,Albrecht:1992kf,Polarski:1995jg}; see also \cite{Aoki:2024bpj,Copinger:2024pai,Fernandes:2024xqr} for a more recent perspective. 
The particle pairs in a squeezed state are produced independently of one another.
This contrasts with coherent states in which single particles are radiated independently of one another.
Coherent single-particle production is basically what we compute to access gravitational waveforms from scattering amplitudes; in a two-particle scattering situation, this means that the gravitational-wave production process is reduced to computing a $2+2+1$ particle amplitude.
Squeezing in a two black hole scattering scenario would involve studying six-particle amplitudes: two for the radiated pairs, plus the four incoming and outgoing black holes. 

\subsection{Bogoliubov $A$ is a generalised amplitude}
\label{sec:bogA}

Scattering amplitudes are matrix elements of the scattering operator, that is objects of the form
\[
\braket{p_1' \cdots p_n' | S | p_1 \cdots p_m} = \braket{\vac | a(p_1') \cdots a(p_n') S a^\dagger(p_1) \cdots a^\dagger(p_m) | \vac} \,.
\]
A more general set of objects has received attention recently~\cite{Kosower:2018adc,Cristofoli:2021vyo,Caron-Huot:2023vxl,Caron-Huot:2023ikn}.
For example, the gravitational waveform~\cite{Cristofoli:2021vyo,Elkhidir:2023dco} involves an integral of the object
\[
\braket{0 | a(p_1') a(p_2') S^\dagger a(k) S a(p_1) a(p_2) | 0} \,,
\]
which can be interpreted as an in-in expectation value.
It is obviously possible to contemplate on-shell bra-kets involving more powers of the $S$ matrix and its inverse.
Remarkably, the usual crossing procedure relates scattering amplitudes to these more general objects~\cite{Caron-Huot:2023vxl,Caron-Huot:2023ikn}.
We shall refer to this class of object as generalised amplitudes.

With this in mind, recall that we defined the Bogoliubov $A$ matrix in equation~\eqref{eq:BcoeffsDef} as an integral over mode functions.
This is reminiscent of the overlap between two single-particle states in non-relativistic quantum mechanics, and immediately suggests that the Bogoliubov coefficients are on-shell quantities of some kind.
Indeed, using equation~\eqref{eq:bGivenBya} we see that
\[ \label{eq:bogAasAmplitude}
\braket{\vac | S^\dagger \, a(k) \, S a^\dagger(p) | \vac} = A(k, p) \,.
\]
Thus $A$ is a generalised amplitude, more precisely an in-in expectation value. 
The same can be said for the Bogoliubov $B$ coefficient, which we will discuss in more detail in section~\ref{sec:BogB}.
These generalised amplitudes are not standard amplitudes\footnote{One could always insert a resolution of the identity inside \eqref{eq:bogAasAmplitude} to relate it  to products of standard amplitudes and their conjugates. This is explored in more detail in reference~\cite{Aoude:2026bek}.}, because $\bra{\vac} S^\dagger$ is not proportional to the vacuum $\bra{\vac}$ as is clear from equation~\eqref{eq:squeezedVacuum}.

The discussion so far has been quite general; now let's see how it applies to the computation of the Hawking ``amplitude'' $\mathcal{A}_H$ in earlier sections of the paper.
We defined $\mathcal{A}_H$ in equation~\eqref{eq:Ahawkfull} as an overlap of wavefunctions precisely following the definition~\eqref{eq:BcoeffsDef} of $A(k, p)$.
So $\mathcal{A}_H$ is a Bogoliubov $A$ coefficient, and indeed this is how it originally appeared in Hawking's work. 
What is interesting for us is that $\mathcal{A}_H$ is a generalised amplitude.
However, this raises a question: we computed the Hawking amplitude using Feynman diagrams, but the diagrams should be a tool for computing amplitudes, not their generalisations.
So what's going on?

We now show that in the present case Feynman diagrams can be used to compute generalised amplitudes, not (just) amplitudes. 
The usual textbook argument relating the Dyson series for $S$ matrix elements to Feynman diagrams via correlators needs care in the presence of backgrounds because we typically make use of the statement\footnote{Compare, for example, equations 4.28 and 4.29 in Peskin and Schroeder~\cite{Peskin:1995ev}.} $\bra{\vac} S^\dagger \propto \bra{\vac}$ which holds in vacuum quantum field theory, but not in the case at hand. With a background switched on, one needs to distinguish between in-in correlators and in-out correlators (which involve the future vacuum $\bra{\Omega, \textrm{out}}$).

To understand the link between Bogoliubov coefficients and diagrams, we focus more specifically on a gravitationally coupled massless scalar field, writing the action as
\[
S_\textrm{action} =\frac{1}2  \int \dd^4 x \left( \eta^{\mu\nu} + h^{\mu\nu} \right) \partial_\mu \phi \, \partial_\nu \phi \,.
\]
We have written $\sqrt{-g} g^{\mu\nu} = \eta^{\mu\nu} + h^{\mu\nu}$; this can be taken as an exact definition of the field $h$.
We assume that the metric is asymptotically Minkowski in the sense that $h^{\mu\nu}(x) \to 0$ sufficiently far in the future and past.

The mode functions are solutions of the interacting Klein-Gordon equation
so we can determine the Bogoliubov coefficients by finding solutions subject to appropriate boundary conditions.
For definiteness, we determine the future mode functions $F(x, k)$.
Our perturbative method is the standard Lippmann-Schwinger iterative series, sometimes known as old-fashioned perturbation theory.
It is simplest to work in Fourier space, where the equation of motion is 
\[\label{eomf}
p^2 \mathsf{F}(p, p_0) =- \int \hat{\dd}^4 q \, p \cdot \sfh(p-q) \cdot q \, \mathsf{F}(q, p_0) \,.
\]
Above, we introduced a convenient metric contraction  
\begin{equation}
p_1\cdot   \sfh(q)\cdot p_2 \equiv \sfh_{\mu\nu}(q)p^{\mu}_1 p_2^{\nu},
\end{equation}
and a serif font notation for Fourier-transformed quantities
\[\label{FTh}
\mathsf{f}^{\mu_{1}\cdots \mu_{n}}(k)\equiv \int \dd^4 x \, e^{ik\cdot x} {f}^{\mu_{1}\cdots \mu_{n}}(x).
\]
Equation \eqref{eomf} has the formal solution (note that $p_0^2 = 0$)
\[
\mathsf{F}(p, p_0) = \hat{\delta}^4(p-p_0) -  \int \hat{\dd}^4 q \, \frac{p \cdot \sfh(p-q) \cdot q}{p^2} \, \mathsf{F}(q, p_0) \,.
\]
Hence the zeroth order solution is 
\[
\mathsf{F}^{(0)} (p, p_0) = \hat{\delta}^4(p-p_0) \,,
\]
and, as can be proven by induction, the remaining terms  in the series for $n \ge 1$ are 
\[\label{eq:perturbativeF}
\mathsf{F}^{(n)}(p_n, p_0) = (-1)^n \int \hat{\dd}^4 p_{n-1} \cdots \hat{\dd}^4 p_{1} \frac{p_n \cdot \sfh(p_n - p_{n-1}) \cdot p_{n-1}}{p_{n}^2} \cdots
\frac{p_1 \cdot \sfh(p_1 - p_0) \cdot p_{0}}{p_{1}^2} \,.
\]
The boundary condition~\eqref{eq:futureBC} requires that the propagator poles are displaced into the upper half-plane so that $p_i^2$ in the denominators should be interpreted as $(p_i^0 - i \epsilon)^2 - {\bm p}_i^2$. 
In position space, the perturbative correction is
\[\label{eq:futureModeIntegral}
F^{(n)}(x, p_0) = (-1)^n \! \int \! \hat{\dd}^4 p_n  \cdots \hat{\dd}^4 p_{1} {e^{-i p_n \cdot x}}  \frac{p_n \cdot \sfh(p_n {-} p_{n-1}) \cdot p_{n-1}}{p_{n}^2} \cdots
\frac{p_1 \cdot \sfh(p_1 {-} p_0) \cdot p_{0}}{p_{1}^2}.
\]
As $t \rightarrow \infty$, we may close the $p_n^0$ contour in the lower half-plane avoiding the poles of the $1 / p_n^2$ propagator. 
The analytic structure of $\sfh$ yields contributions, but these are suppressed as $t \rightarrow \infty$ because of our hypothesis that the metric has finite spacetime support.

We now deduce a perturbative series for $A(p, k)$ using its definition~\eqref{eq:BcoeffsDef}. 
Since the inner product is time-independent, we evaluate the integral in equation~\eqref{eq:BcoeffsDef} as $t \rightarrow - \infty$, so that the past mode functions are simple exponentials.
We can therefore take advantage of the fact~\eqref{eq:positiveFreqPart} that the inner product projects onto the positive-frequency Fourier component of $F(x, k)$.
The problem is reduced to determining this positive-frequency part using equation~\eqref{eq:futureModeIntegral}.
As $t \rightarrow - \infty$, we close the $p_n^0$ contour of integration in the upper half-plane.
Now both the explicit propagator poles contribute as well as terms from $\sfh$. 
However the latter are suppressed as $t \rightarrow -\infty$ (again, because of the finite support of $h_{\mu\nu}(x)$).
The result is 
\[
A^*(p_n, p_0) = i(-1)^n\int & \hat{\dd}^4 p_{n-1}  \cdots \hat{\dd}^4 p_{1} \, 
p_n \cdot \sfh(p_n - p_{n-1}) \cdot p_{n-1} \\
&\times \frac{p_{n-1} \cdot \sfh(p_{n-1} - p_{n-2}) \cdot p_{n-2}}{p_{n-1}^2} \cdots \frac{p_1 \cdot \sfh(p_1 - p_0) \cdot p_{0} }{p_{1}^2}\,.
\]
The expression~\eqref{eq:perturbativeF} has a clear diagrammatic interpretation, illustrated here for $n=3$:
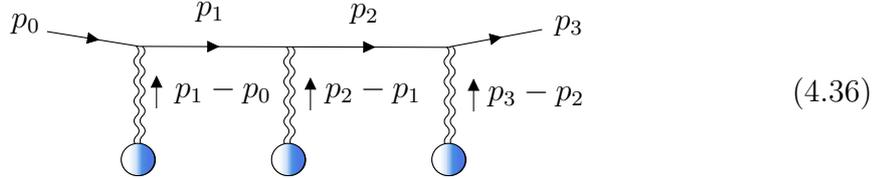
\[  
\begin{tikzpicture}[x=0.75pt,y=0.75pt,yscale=-1,xscale=1]

  
\tikzset {_0wji2jnyu/.code = {\pgfsetadditionalshadetransform{ \pgftransformshift{\pgfpoint{0 bp } { 0 bp }  }  \pgftransformrotate{0 }  \pgftransformscale{2 }  }}}
\pgfdeclarehorizontalshading{_e4pmy6i3w}{150bp}{rgb(0bp)=(1,1,1);
rgb(42.606724330357146bp)=(1,1,1);
rgb(45.89285714285714bp)=(0.82,0.89,0.97);
rgb(57.14285714285714bp)=(0.29,0.56,0.89);
rgb(57.25bp)=(0.29,0.56,0.89);
rgb(62.5bp)=(0.29,0.56,0.89);
rgb(100bp)=(0.29,0.56,0.89)}

  
\tikzset {_g3ermf92b/.code = {\pgfsetadditionalshadetransform{ \pgftransformshift{\pgfpoint{0 bp } { 0 bp }  }  \pgftransformrotate{0 }  \pgftransformscale{2 }  }}}
\pgfdeclarehorizontalshading{_jwf0qzyp2}{150bp}{rgb(0bp)=(1,1,1);
rgb(42.428152901785715bp)=(1,1,1);
rgb(47.67857142857143bp)=(0.82,0.89,0.97);
rgb(52.23214285714286bp)=(0.29,0.56,0.89);
rgb(57.5bp)=(0.29,0.47,0.89);
rgb(62.42815290178571bp)=(0.29,0.47,0.89);
rgb(100bp)=(0.29,0.47,0.89)}

  
\tikzset {_3fmdg3rbr/.code = {\pgfsetadditionalshadetransform{ \pgftransformshift{\pgfpoint{0 bp } { 0 bp }  }  \pgftransformrotate{0 }  \pgftransformscale{2 }  }}}
\pgfdeclarehorizontalshading{_cfo78hfis}{150bp}{rgb(0bp)=(1,1,1);
rgb(42.606724330357146bp)=(1,1,1);
rgb(45.89285714285714bp)=(0.82,0.89,0.97);
rgb(57.14285714285714bp)=(0.29,0.56,0.89);
rgb(57.25bp)=(0.29,0.56,0.89);
rgb(62.5bp)=(0.29,0.56,0.89);
rgb(100bp)=(0.29,0.56,0.89)}

  
\tikzset {_ydgfqjrcm/.code = {\pgfsetadditionalshadetransform{ \pgftransformshift{\pgfpoint{0 bp } { 0 bp }  }  \pgftransformrotate{0 }  \pgftransformscale{2 }  }}}
\pgfdeclarehorizontalshading{_k3ztdvr5x}{150bp}{rgb(0bp)=(1,1,1);
rgb(42.428152901785715bp)=(1,1,1);
rgb(47.67857142857143bp)=(0.82,0.89,0.97);
rgb(52.23214285714286bp)=(0.29,0.56,0.89);
rgb(57.5bp)=(0.29,0.47,0.89);
rgb(62.42815290178571bp)=(0.29,0.47,0.89);
rgb(100bp)=(0.29,0.47,0.89)}

  
\tikzset {_n5uagg2as/.code = {\pgfsetadditionalshadetransform{ \pgftransformshift{\pgfpoint{0 bp } { 0 bp }  }  \pgftransformrotate{0 }  \pgftransformscale{2 }  }}}
\pgfdeclarehorizontalshading{_5myturyio}{150bp}{rgb(0bp)=(1,1,1);
rgb(42.606724330357146bp)=(1,1,1);
rgb(45.89285714285714bp)=(0.82,0.89,0.97);
rgb(57.14285714285714bp)=(0.29,0.56,0.89);
rgb(57.25bp)=(0.29,0.56,0.89);
rgb(62.5bp)=(0.29,0.56,0.89);
rgb(100bp)=(0.29,0.56,0.89)}

  
\tikzset {_gvxzsfs9w/.code = {\pgfsetadditionalshadetransform{ \pgftransformshift{\pgfpoint{0 bp } { 0 bp }  }  \pgftransformrotate{0 }  \pgftransformscale{2 }  }}}
\pgfdeclarehorizontalshading{_13ej6v826}{150bp}{rgb(0bp)=(1,1,1);
rgb(42.428152901785715bp)=(1,1,1);
rgb(47.67857142857143bp)=(0.82,0.89,0.97);
rgb(52.23214285714286bp)=(0.29,0.56,0.89);
rgb(57.5bp)=(0.29,0.47,0.89);
rgb(62.42815290178571bp)=(0.29,0.47,0.89);
rgb(100bp)=(0.29,0.47,0.89)}
\tikzset{every picture/.style={line width=0.75pt}} 


\draw    (96,16929) -- (142.02,16936.46) ;
\draw [shift={(121.58,16933.15)}, rotate = 189.21] [fill={rgb, 255:red, 0; green, 0; blue, 0 }  ][line width=0.08]  [draw opacity=0] (5.36,-2.57) -- (0,0) -- (5.36,2.57) -- cycle    ;
\draw    (296.82,16936.69) -- (343,16928) ;
\draw [shift={(322.47,16931.87)}, rotate = 169.34] [fill={rgb, 255:red, 0; green, 0; blue, 0 }  ][line width=0.08]  [draw opacity=0] (5.36,-2.57) -- (0,0) -- (5.36,2.57) -- cycle    ;
\draw    (142.02,16936.46) -- (215.82,16936.69) ;
\draw [shift={(181.52,16936.58)}, rotate = 180.18] [fill={rgb, 255:red, 0; green, 0; blue, 0 }  ][line width=0.08]  [draw opacity=0] (5.36,-2.57) -- (0,0) -- (5.36,2.57) -- cycle    ;
\draw    (215.82,16936.69) -- (297.63,16936.93) ;
\draw [shift={(259.33,16936.82)}, rotate = 180.16] [fill={rgb, 255:red, 0; green, 0; blue, 0 }  ][line width=0.08]  [draw opacity=0] (5.36,-2.57) -- (0,0) -- (5.36,2.57) -- cycle    ;
\draw    (150.6,16954.62) -- (150.6,16967.62) ;
\draw [shift={(150.6,16951.62)}, rotate = 90] [fill={rgb, 255:red, 0; green, 0; blue, 0 }  ][line width=0.08]  [draw opacity=0] (5.36,-2.57) -- (0,0) -- (5.36,2.57) -- cycle    ;
\draw    (227.4,16955.62) -- (227.4,16968.62) ;
\draw [shift={(227.4,16952.62)}, rotate = 90] [fill={rgb, 255:red, 0; green, 0; blue, 0 }  ][line width=0.08]  [draw opacity=0] (5.36,-2.57) -- (0,0) -- (5.36,2.57) -- cycle    ;
\draw    (298.32,16936.69) .. controls (299.99,16938.35) and (300,16940.02) .. (298.34,16941.69) .. controls (296.68,16943.36) and (296.68,16945.02) .. (298.35,16946.69) .. controls (300.02,16948.35) and (300.03,16950.02) .. (298.37,16951.69) .. controls (296.71,16953.36) and (296.71,16955.02) .. (298.38,16956.69) .. controls (300.05,16958.35) and (300.06,16960.02) .. (298.4,16961.69) .. controls (296.74,16963.36) and (296.74,16965.02) .. (298.41,16966.69) .. controls (300.08,16968.36) and (300.08,16970.02) .. (298.42,16971.69) .. controls (296.76,16973.36) and (296.77,16975.03) .. (298.44,16976.69) .. controls (300.11,16978.36) and (300.11,16980.02) .. (298.45,16981.69) .. controls (296.79,16983.36) and (296.8,16985.03) .. (298.47,16986.69) -- (298.48,16989.91) -- (298.48,16989.91)(295.32,16936.7) .. controls (296.99,16938.36) and (297,16940.03) .. (295.34,16941.7) .. controls (293.68,16943.37) and (293.68,16945.03) .. (295.35,16946.7) .. controls (297.02,16948.36) and (297.03,16950.03) .. (295.37,16951.7) .. controls (293.71,16953.37) and (293.71,16955.03) .. (295.38,16956.7) .. controls (297.05,16958.36) and (297.06,16960.03) .. (295.4,16961.7) .. controls (293.74,16963.37) and (293.74,16965.03) .. (295.41,16966.7) .. controls (297.08,16968.37) and (297.08,16970.03) .. (295.42,16971.7) .. controls (293.76,16973.37) and (293.77,16975.04) .. (295.44,16976.7) .. controls (297.11,16978.37) and (297.11,16980.03) .. (295.45,16981.7) .. controls (293.79,16983.37) and (293.8,16985.04) .. (295.47,16986.7) -- (295.48,16989.92) -- (295.48,16989.92) ;
\draw    (308.8,16956.22) -- (308.8,16969.22) ;
\draw [shift={(308.8,16953.22)}, rotate = 90] [fill={rgb, 255:red, 0; green, 0; blue, 0 }  ][line width=0.08]  [draw opacity=0] (5.36,-2.57) -- (0,0) -- (5.36,2.57) -- cycle    ;
\path  [shading=_e4pmy6i3w,_0wji2jnyu] (296.58,16985.71) .. controls (291.93,16985.71) and (288.16,16989.23) .. (288.16,16993.58) .. controls (288.16,16997.93) and (291.93,17001.46) .. (296.58,17001.46) .. controls (301.23,17001.46) and (305,16997.93) .. (305,16993.58) .. controls (305,16989.23) and (301.23,16985.71) .. (296.58,16985.71) -- cycle ; 
 \draw   (296.58,16985.71) .. controls (291.93,16985.71) and (288.16,16989.23) .. (288.16,16993.58) .. controls (288.16,16997.93) and (291.93,17001.46) .. (296.58,17001.46) .. controls (301.23,17001.46) and (305,16997.93) .. (305,16993.58) .. controls (305,16989.23) and (301.23,16985.71) .. (296.58,16985.71) -- cycle ; 

\path  [shading=_jwf0qzyp2,_g3ermf92b] (296.42,16985.25) .. controls (301.04,16985.27) and (304.77,16988.93) .. (304.74,16993.42) .. controls (304.72,16997.9) and (300.95,17001.52) .. (296.32,17001.5) .. controls (291.7,17001.48) and (287.97,16997.82) .. (288,16993.33) .. controls (288.03,16988.84) and (291.8,16985.22) .. (296.42,16985.25) -- cycle ; 
 \draw   (296.42,16985.25) .. controls (301.04,16985.27) and (304.77,16988.93) .. (304.74,16993.42) .. controls (304.72,16997.9) and (300.95,17001.52) .. (296.32,17001.5) .. controls (291.7,17001.48) and (287.97,16997.82) .. (288,16993.33) .. controls (288.03,16988.84) and (291.8,16985.22) .. (296.42,16985.25) -- cycle ; 

\draw    (218.32,16936.69) .. controls (219.99,16938.35) and (220,16940.02) .. (218.34,16941.69) .. controls (216.68,16943.36) and (216.68,16945.02) .. (218.35,16946.69) .. controls (220.02,16948.35) and (220.03,16950.02) .. (218.37,16951.69) .. controls (216.71,16953.36) and (216.71,16955.02) .. (218.38,16956.69) .. controls (220.05,16958.35) and (220.06,16960.02) .. (218.4,16961.69) .. controls (216.74,16963.36) and (216.74,16965.02) .. (218.41,16966.69) .. controls (220.08,16968.36) and (220.08,16970.02) .. (218.42,16971.69) .. controls (216.76,16973.36) and (216.77,16975.03) .. (218.44,16976.69) .. controls (220.11,16978.36) and (220.11,16980.02) .. (218.45,16981.69) .. controls (216.79,16983.36) and (216.8,16985.03) .. (218.47,16986.69) -- (218.48,16989.91) -- (218.48,16989.91)(215.32,16936.7) .. controls (216.99,16938.36) and (217,16940.03) .. (215.34,16941.7) .. controls (213.68,16943.37) and (213.68,16945.03) .. (215.35,16946.7) .. controls (217.02,16948.36) and (217.03,16950.03) .. (215.37,16951.7) .. controls (213.71,16953.37) and (213.71,16955.03) .. (215.38,16956.7) .. controls (217.05,16958.36) and (217.06,16960.03) .. (215.4,16961.7) .. controls (213.74,16963.37) and (213.74,16965.03) .. (215.41,16966.7) .. controls (217.08,16968.37) and (217.08,16970.03) .. (215.42,16971.7) .. controls (213.76,16973.37) and (213.77,16975.04) .. (215.44,16976.7) .. controls (217.11,16978.37) and (217.11,16980.03) .. (215.45,16981.7) .. controls (213.79,16983.37) and (213.8,16985.04) .. (215.47,16986.7) -- (215.48,16989.92) -- (215.48,16989.92) ;
\path  [shading=_cfo78hfis,_3fmdg3rbr] (216.58,16985.71) .. controls (211.93,16985.71) and (208.16,16989.23) .. (208.16,16993.58) .. controls (208.16,16997.93) and (211.93,17001.46) .. (216.58,17001.46) .. controls (221.23,17001.46) and (225,16997.93) .. (225,16993.58) .. controls (225,16989.23) and (221.23,16985.71) .. (216.58,16985.71) -- cycle ; 
 \draw   (216.58,16985.71) .. controls (211.93,16985.71) and (208.16,16989.23) .. (208.16,16993.58) .. controls (208.16,16997.93) and (211.93,17001.46) .. (216.58,17001.46) .. controls (221.23,17001.46) and (225,16997.93) .. (225,16993.58) .. controls (225,16989.23) and (221.23,16985.71) .. (216.58,16985.71) -- cycle ; 

\path  [shading=_k3ztdvr5x,_ydgfqjrcm] (216.42,16985.25) .. controls (221.04,16985.27) and (224.77,16988.93) .. (224.74,16993.42) .. controls (224.72,16997.9) and (220.95,17001.52) .. (216.32,17001.5) .. controls (211.7,17001.48) and (207.97,16997.82) .. (208,16993.33) .. controls (208.03,16988.84) and (211.8,16985.22) .. (216.42,16985.25) -- cycle ; 
 \draw   (216.42,16985.25) .. controls (221.04,16985.27) and (224.77,16988.93) .. (224.74,16993.42) .. controls (224.72,16997.9) and (220.95,17001.52) .. (216.32,17001.5) .. controls (211.7,17001.48) and (207.97,16997.82) .. (208,16993.33) .. controls (208.03,16988.84) and (211.8,16985.22) .. (216.42,16985.25) -- cycle ; 

\draw    (143.52,16936.45) .. controls (145.19,16938.12) and (145.19,16939.78) .. (143.53,16941.45) .. controls (141.87,16943.12) and (141.88,16944.79) .. (143.55,16946.45) .. controls (145.22,16948.12) and (145.22,16949.78) .. (143.56,16951.45) .. controls (141.9,16953.12) and (141.91,16954.79) .. (143.58,16956.45) .. controls (145.25,16958.12) and (145.25,16959.78) .. (143.59,16961.45) .. controls (141.93,16963.12) and (141.93,16964.78) .. (143.6,16966.45) .. controls (145.27,16968.11) and (145.28,16969.78) .. (143.62,16971.45) .. controls (141.96,16973.12) and (141.96,16974.78) .. (143.63,16976.45) .. controls (145.3,16978.11) and (145.31,16979.78) .. (143.65,16981.45) .. controls (141.99,16983.12) and (141.99,16984.78) .. (143.66,16986.45) -- (143.67,16989.67) -- (143.67,16989.67)(140.52,16936.46) .. controls (142.19,16938.13) and (142.19,16939.79) .. (140.53,16941.46) .. controls (138.87,16943.13) and (138.88,16944.8) .. (140.55,16946.46) .. controls (142.22,16948.13) and (142.22,16949.79) .. (140.56,16951.46) .. controls (138.9,16953.13) and (138.91,16954.8) .. (140.58,16956.46) .. controls (142.25,16958.13) and (142.25,16959.79) .. (140.59,16961.46) .. controls (138.93,16963.13) and (138.93,16964.79) .. (140.6,16966.46) .. controls (142.27,16968.12) and (142.28,16969.79) .. (140.62,16971.46) .. controls (138.96,16973.13) and (138.96,16974.79) .. (140.63,16976.46) .. controls (142.3,16978.12) and (142.31,16979.79) .. (140.65,16981.46) .. controls (138.99,16983.13) and (138.99,16984.79) .. (140.66,16986.46) -- (140.67,16989.68) -- (140.67,16989.68) ;
\path  [shading=_5myturyio,_n5uagg2as] (141.58,16985.71) .. controls (136.93,16985.71) and (133.16,16989.23) .. (133.16,16993.58) .. controls (133.16,16997.93) and (136.93,17001.46) .. (141.58,17001.46) .. controls (146.23,17001.46) and (150,16997.93) .. (150,16993.58) .. controls (150,16989.23) and (146.23,16985.71) .. (141.58,16985.71) -- cycle ; 
 \draw   (141.58,16985.71) .. controls (136.93,16985.71) and (133.16,16989.23) .. (133.16,16993.58) .. controls (133.16,16997.93) and (136.93,17001.46) .. (141.58,17001.46) .. controls (146.23,17001.46) and (150,16997.93) .. (150,16993.58) .. controls (150,16989.23) and (146.23,16985.71) .. (141.58,16985.71) -- cycle ; 

\path  [shading=_13ej6v826,_gvxzsfs9w] (141.42,16985.25) .. controls (146.04,16985.27) and (149.77,16988.93) .. (149.74,16993.42) .. controls (149.72,16997.9) and (145.95,17001.52) .. (141.32,17001.5) .. controls (136.7,17001.48) and (132.97,16997.82) .. (133,16993.33) .. controls (133.03,16988.84) and (136.8,16985.22) .. (141.42,16985.25) -- cycle ; 
 \draw   (141.42,16985.25) .. controls (146.04,16985.27) and (149.77,16988.93) .. (149.74,16993.42) .. controls (149.72,16997.9) and (145.95,17001.52) .. (141.32,17001.5) .. controls (136.7,17001.48) and (132.97,16997.82) .. (133,16993.33) .. controls (133.03,16988.84) and (136.8,16985.22) .. (141.42,16985.25) -- cycle ; 

\draw (77,16919) node [anchor=north west][inner sep=0.75pt]    {$p_{0}$};
\draw (348,16920) node [anchor=north west][inner sep=0.75pt]    {$p_{3}$};
\draw (158.6,16951.62) node [anchor=north west][inner sep=0.75pt]    {$p_{1} -p_{0}$};
\draw (169,16913) node [anchor=north west][inner sep=0.75pt]    {$p_{1}$};
\draw (246,16914) node [anchor=north west][inner sep=0.75pt]    {$p_{2}$};
\draw (233.4,16951.62) node [anchor=north west][inner sep=0.75pt]    {$p_{2} -p_{1}$};
\draw (314.8,16953.22) node [anchor=north west][inner sep=0.75pt]    {$p_{3} -p_{2}$};

\end{tikzpicture}
\]
These are precisely the diagrams which were studied and resummed in section~\ref{sec:Exp_Leading}. 
Thus our diagrammatic approach was appropriate for the computation of a particular generalised amplitude: the Bogoliubov $A$ coefficient\footnote{A comment on the $i\epsilon$ prescription in section~\ref{sec:Exp_Leading}. As the energy of our incoming particle $p^0$ is large in the geometric optics expansion, the difference between the retarded propagator and the Feynman propagator is subdominant in our computation.}.

\subsection{Forwards or backwards evolution?}
\label{sec:forwardsVbackwards}

In the scattering amplitudes computations of section~\ref{sec:Tree_level} we evolved an initial state in the far past forwards in time.
However our Lippmann-Schwinger computations, following more closely Hawking's setup, express a future single-particle wavefunction in terms of a sum of past wavefunctions. 
In other words we evolved backwards in time.
Now let us compare forward and backward evolution.

The simplest starting point is to return to the ray tracing computation, but now evolving forwards in time.
We take an initial state as $t\rightarrow -\infty$
\[
P(v, E_0) = \frac{1}{2 E_0 r} e^{-i E_0 v} \,,
\]
defined on $\scri^-$.
Following precisely the same ray-tracing logic as in section~\ref{sec:Review} we know that corresponding solution in the future ($t \rightarrow \infty$) is
\[
P(u, E_0) = \frac{1}{2 E_0 r} e^{-i E_0 v(u)} \,,
\]
where $u(v)$ is determined by the ray-tracing relation~\eqref{eq:rayTracingRelation}
\[
u &= v(u) - 4 GM \log (v_0 - v(u))/\mu \,.
\]
Decomposing this solution in terms of outgoing future modes 
\[
F(u, E) = \frac{1}{2 Er} e^{-i Eu}
\]
using straightforward Fourier analysis, we see that\footnote{The measure $\textrm{d} E \, E/ 4\pi^2$ is appropriate for spherical wavefunctions in our conventions, see appendix~\ref{app:sphericalNorms}.}
\[
P(u, E_0) = \int_{- \infty}^{\infty} \frac{\textrm{d}E \, E}{4\pi^2} F(u, E) \left[\frac{2\pi}{E_0} \int \mathrm{d} u \, e^{i Eu} e^{-i E_0 v(u)} \right] \,.
\]
Evidently the amplitude is the overlap between $P$ and $F$ for positive energy $E$. 
That is,
\[
\mathcal{A} = \frac{2\pi}{E_0} \int \mathrm{d} u \, e^{i Eu} e^{-i E_0 v(u)} \,.
\]
This can be simplified by changing variable of integration to $v$ rather than $u$ so that 
\[\label{eq:Aagain}
\mathcal{A} = \frac{2\pi}{E_0} \int \mathrm{d} v \left(1 - \frac{4GM}v + \cdots \right) e^{iE (v - 4GM \log( -v)/ \mu)} e^{-i E_0 v} \,.
\]
The quantity in parentheses above arises from the Jacobian;
the leading term is simply unity, and
we will meet the subleading term $- 4 GM / v$ again in the next section, where it will arise from subleading/loop corrections.
The dominant term in equation~\eqref{eq:Aagain} matches the amplitude of equation~\eqref{eq:aHgeneraltree} in detail.

Notice that the amplitude $\mathcal{A}$ differs from the backwards evolution $\mathcal{A}_H$ in equation~\eqref{eq:AHawking} by a factor $E/E_0$ (this distinction motivated our notation $\mathcal{A}_H$ and $\mathcal{A}$).
That is,
\[ \label{eq:energyRatio}
\mathcal{A}_H = \frac {E_0}E \mathcal{A} \,.
\]
We included this factor in our discussion of the Hawking spectrum in equation~\eqref{eq:dnDistn}.

It is possible to repeat the scattering amplitudes computation of section~\ref{sec:Tree_level}, now evolving backwards in time, to recover the same factor. 
In doing so, one must take care in the energy approximation corresponding to equation~\eqref{der}.
The key point is that in this approximation one should retain the energy of the state which is later in time.
We chose to present our scattering amplitudes computation in the usual time order because this is more familiar,\footnote{ We believe it would be interesting to investigate how this story interplays with the findings of \cite{Donoghue:2019ecz}.} and because our main interest is in the phase of the amplitude.

\subsection{Bogoliubov $B$ as crossing}
\label{sec:BogB}

Returning to equation~\eqref{eq:bGivenBya}, we see that the Bogoliubov $B$ coefficient is also a generalised amplitude, namely
\[
\braket{\vac | a(p) \, S^\dagger \, a(k) \, S | \vac } = B(k, p) \,.
\]
It is interesting to compare with equation~\eqref{eq:bogAasAmplitude} for the $A$ coefficient.
The expressions for $A$ and $B$ are reminiscent of crossing: starting with $A$, one obtains $B$ by ``crossing'' the creation operator $a^\dagger(p)$ into an annihilation operator $a(p)$.
This crossing can be understood as an analytic continuation of integrals when $A$ and $B$ are expressed in terms of mode functions.
Writing the Klein-Gordon inner product in the far past, we have 
\[\label{eq:PWbogs}
A(k, p) &= \int \dd^3 x \left( \bar{\mathsf{F}}(k) i \partial_t e^{-i p \cdot x } - e^{-i p \cdot x} i \partial_t \bar{\mathsf{F}}(k) \right) \,, \\
B(k, p) &= \int \dd^3 x \left( \bar{\mathsf{F}}(k) i \partial_t e^{i p \cdot x } - e^{i p \cdot x} i \partial_t \bar{\mathsf{F}}(k) \right) \,.
\]
Thus crossing is an analytic continuation from $p$ to $-p$ in the integrals\footnote{This crossing has recently been studied in more detail, including at the level of diagrams, in reference~\cite{Aoude:2026bek}.}. 
This analytic continuation is well known, and Hawking used precisely this link in his original calculation~\cite{Hawking:1975vcx}.

More concretely, we wish to apply this crossing to our amplitude~\eqref{eq:aHspherical} which is written in terms of spherical ($\mathcal{\ell} = 0, m = 0$) states.
We can obtain expressions for the Bogoliubov coefficients in terms of these states by angular averaging equation~\eqref{eq:PWbogs}.
Alternatively we may proceed directly using Fourier analysis; since the interactions are spherically symmetric, we need only relate spherical states to spherical states, which reduces the problem to integrals over a single Fourier variable.
A basis of spherical states in the far past is
\[
P(v, E_0) = \frac{1}{2 E_0 r} e^{-i E_0 v} \,,
\]
so we may write the wavefunction of a future single-particle spherical eigenstate as
\[
F(v, E) = \int_0^\infty \frac{E_0 \textrm{d}E_0}{4\pi^2} \left( P(v, E_0) A(E, E_0) + \bar{P}(v, E_0) B(E, E_0) \right) \,,
\]
where Fourier analysis implies that the Bogoliubov coefficients are
\[
A(E, E_0) = 8 \pi \int \textrm{d} v \, r^2 F(v, E) (-i \partial_v) \bar P(E_0, v) \\
B(E, E_0) = 8 \pi \int \textrm{d} v \, r^2 F(v, E) (i \partial_v) P(E_0, v) .
\]
Because $A(E, E_0)$ is the Hawking amplitude $\mathcal{A}$ given in equation~\eqref{eq:aHspherical}, we see that the Bogoliubov $B$ coefficient is obtained by analytically continuing $E_0 \rightarrow -E_0$ with the result
\[
\mathcal{B} = \frac{2\pi}{E_0} \int_{-\infty}^0 \dd v \, e^{i(E+E_0) v}
    e^{-4iGM E\log(-v/\mu)} \,.
\]
We ignored an irrelevant sign, and chose a notation to emphasise the link to the amplitude $\mathcal{A}$ which we obtained by summing a series of diagrams.

Now let us turn to physical observables.
An important example is the number of particles of momentum $p$ created from the vacuum.
In terms of plane-wave states, the number is
\[
n &= \int \dd \Phi(p) \braket{\vac| S^\dagger a^\dagger(p) a(p) S | \vac}  \\
&= \int \dd \Phi(p, k) \, B^*(p, k) B(p, k) \,.
\] 
This illustrates the importance of the $B$ coefficient: it contains information about the number distribution of emitted particles.
In view of the double phase space integration, we may define a differential number distribution 
\[
\dd n = \dd\Phi(p, k) \, B^*(p, k) B(p, k) \,.
\]
In terms of spherical states this distribution is
\[
\dd n = \frac{E \dd E}{4\pi^2} \frac{E_0 \dd E_0}{4\pi^2} 
|\mathcal{B}|^2 \left(\frac{E_0}{E}\right)^2 \,,
\]
including the energy ratio from equation~\eqref{eq:energyRatio}.
Following the discussion of section~\ref{spectrum} but now working with $\mathcal{B}$, which is   the analytic continuation of $\mathcal{A}$, one finds
\[
\dd n = \dd E \, \dd E_0 \frac{2 G M}{\pi E_0} \frac{1}{e^{8\pi GM E}-1} \, .
\]
The result is a thermal distribution, now without the additional factor in the numerator.

Before we move on, let us recapitulate our overall logic.
Up to normalisation, we have seen that the Bogoliubov $A$ coefficient can be obtained by resumming a series of diagrams. 
The Bogoliubov $B$ coefficient captures the number distribution of emitted particles, and it can be determined by crossing the $A$ coefficient. 
In this way we can access the thermal distribution from Feynman diagrams.

There is (at least) one puzzling aspect of this connection.
Classically, we know that the sum of diagrams in $\mathcal{A}$ is related to a well-defined classical trajectory: a lightlike geodesic in the Vaidya background.
However after crossing the interpretation of this trajectory is much less clear. 
It is tempting to suppose that the classical trajectory goes over to a semiclassical WKB-style tunnelling solution.
If true, this would make contact with the approach of reference~\cite{Parikh:1999mf}.

Now that we have discussed the connection between scattering amplitudes and Bogoliubov coefficients, we return to perturbative calculations. 
We will confirm the subleading correction to the eikonal phase predicted by equation~\eqref{eq:AHawking} using   one-loop amplitudes.

\section{NLO eikonal phase and horizon contribution}
\label{sec:OneLoop}
In section~\ref{sec:Tree_level} we showed that an infinite sum over (leading) loop orders yields an exponential which reproduces Hawking's ray-tracing relation~\eqref{eq:rayTracingRelation}, and hence (after analytic continuation) the Hawking radiation. 
In this section we will dig a bit deeper into the subleading part of the one loop diagram.
We will see that this computes the 
order $G^2M^2$ correction to the phase in equation~\eqref{eq:AHawking}.
This correction arises from expanding $\log (v_0 - v)$ in the full Hawking amplitude~\eqref{eq:Ahawkfull} as a series in $v_0/v$.
Recall that $v_0 = -4GM$ is the advanced time corresponding to the horizon of the black hole, which is a finite distance from $v = 0$ because the black hole has a finite radius (it may help to review the discussion above equation~\eqref{eq:epsilon}).
We therefore interpret these corrections as finite size effects:
they are directly sensitive to the scale of the event horizon.

Before turning to the one loop problem, we begin by returning to the tree-level case.
Although our interest is restricted to the leading order of the geometric optics limit --- this is where Hawking's ray-tracing relation holds ---  we will actually need to determine the subleading piece of the tree amplitude in the geometric-optics limit. 
This occurs because we need to distinguish this term from the leading-in-geometric optics, but subleading in $GM$, parts which we wish to compute.
The same phenomenon occurs in the eikonal approximation, albeit at one higher order in perturbation theory~\cite{DiVecchia:2023frv}.

\subsection{Review of the tree-level eikonal and of its subleading corrections}

Let us begin by repeating the tree-level exercise using a more covariant notation that will be extremely convenient at one loop.  Indeed, one can write the interaction Lagrangian  as 
\begin{align}
\mathcal{L}_{\rm int.} = \frac{2GM(\sfk\cdot x)}{r} (\sfk\cdot \partial \phi)^2 = \partial_\mu \phi\, h^{\mu\nu}(x)\,\partial_\nu \phi,
\end{align}
where for the sake of generality we momentarily revert to an unspecified  mass function $M(\sfk\cdot x)$.

The tree-level amplitude in momentum space is
\begin{align}
i\cA_{0}(q) \equiv  i \int \dd^4 x\, e^{iq\cdot x}\,p_\mu\, h^{\mu\nu}(x) (p-q)_{\nu} \,.
\end{align}
The leading contribution, proportional to $p_\mu p_\nu$, was evaluated in section~\ref{sec:Tree_level}. 
The  term above proportional to $h^{\mu\nu}p_\mu q_\nu$ is  a new  contribution with subleading scaling in the geometric-optics limit. 
It will be convenient to work in position space, where we obtain
\begin{align}
\label{eq:Covariant_Treelvl_Amp}
i\tilde{\mathcal{A}}_0(b) 
\equiv 
i\int \dd\lambda\, \hat{\dd}^4 q\, \dd^4x\, e^{i(x-b-2\lambda\,p)\cdot q+i\lambda q^2}\, p_\mu h^{\mu\nu}(x) (p-q)_{\nu}.
\end{align}
This expression retains the full on-shell condition in contrast to our earlier approximation~\eqref{eq:deltaApprox}.
We see that there is a subleading contribution coming from the expansion of $q^2$ terms above as
 \[\label{qsqq}
 \exp[i\lambda q^2] \simeq 1+ i\lambda q^2.
 \]
Thus, we can  write the tree-level fragments    as
\begin{subequations}
\label{eq:TreeLevel_Leading_Subleading}
\begin{align}
    i\tilde{\mathcal{A}}_0^{(0)}(b) &=  
    i\int \dd\lambda\,\hat{\dd}^4q\,e^{-i(b+2\lambda\,p)\cdot q}
    \,\,\sfh^{\mu\nu}(q)\, p_\mu p_\nu,\\
\label{subeq:interesting}    i\tilde{\mathcal{A}}_0^{(1)}(b) &=  
    i\int \dd\lambda\,\hat{\dd}^4q\,e^{-i(b+2\lambda\,p )\cdot q}
    \,\,\sfh^{\mu\nu}(q)\, \left(i\lambda\, q^2 p_\mu p_\nu-p_\mu q_\nu\right).
\end{align}
\end{subequations}
The first line in equation \eqref{eq:TreeLevel_Leading_Subleading} describes the fragment studied in  section~\ref{sec:Tree_level}, computed implicitly in equation \ref{eq:intlamtree}. We report it again here, written in a covariant form  which will be useful later
\[\label{frag0}
i\tilde{\mathcal{A}}_0^{(0)}(b)=i4GM u\cdot p\log\left(\frac{4u\cdot  p\, \lambda_{\infty}}{-u\cdot b} \right).
\]

Next, we focus on the new $\tilde{\mathcal{A}}_0^{(1)}$ fragment.
To deal with the $q^\mu$ factors in the numerators of equation~\eqref{subeq:interesting} we use classic integration by parts (IBP) tricks  and exchange the momenta with derivatives with respect to $x$. 
In turn, these  lead to expressions involving derivatives of the metric, such as
\begin{align}
\label{eq:Derivative_h}
\partial^\lambda h^{\mu\nu} = G\bigg[\partial^\lambda\left(\frac{\sfk^\mu \sfk^\nu}{r}\right) M(\sfk \cdot x) +   \left(\frac{\sfk^\lambda\sfk^\mu \sfk^\nu}{r}\right)M'(\sfk \cdot x)\bigg],
\end{align}
where we used the chain rule to handle the derivative of $M$
\begin{equation}
    \partial^\lambda M(\sfk \cdot x)   = 
\partial^\lambda(\sfk\cdot x)M'(\sfk \cdot x)=\sfk^\lambda M'(\sfk \cdot x).
\end{equation}
The first term in \eqref{eq:Derivative_h} is familiar from the usual case of the stationary Schwarzschild metric in KS coordinates, while the second term is more interesting as it involves the derivative of the mass distribution; 
in our case with $M(\sfk \cdot x)=M \Theta(\sfk \cdot x)$ we simply have $M'=M \delta(\sfk\cdot x)$. 
Similarly, a second derivative appears stemming from  the $q^2$ term in equation~\eqref{subeq:interesting}. 
Using the fact that $\sfk^2=0$, we find after some straightforward steps that
\begin{align}\label{quantrein}
i\tilde{\mathcal{A}}_0^{(1)}(b) &= 
\int \dd \lambda \frac{GM(\sfk\cdot x)}{u\cdot p}\Theta(\lambda)\frac{1}{\lambda^2} = -\frac{4GM}{u\cdot b} \,.
\end{align}
We used $M(\sfk\cdot x)=M\Theta(\sfk\cdot x)$ for the explicit integration. As discussed above in the context of equation~\eqref{eq:intlamtree}, the Heaviside functions set the limit of integration for the affine parameter $\lambda$.
Covariantly, this is given by
\begin{align}
\Theta(\sfk\cdot x) \neq 0 \quad \Rightarrow \quad  u\cdot b +2u\cdot p\, (\lambda+|\lambda|) \geq 0
\quad\Rightarrow\quad 
\lambda \geq -\frac{u\cdot b}{4u\cdot p}.
\end{align}
In this expression, $u^\mu$ is an arbitrary time direction with $u^2=1$; this can be set to $(1,\bm{0})$ to recover the results of the previous sections. 
It may be worth noting that the term in equation~\eqref{quantrein} already appeared in our ray-tracing discussion: it is the next-to-leading order term in the Jacobian determinant in equation~\eqref{eq:Aagain}, after writing the future affine parameter $u$ in terms of the past $v$.
So we see again that our diagrams correspond to the ray-tracing story.
A second comment on equation~\eqref{quantrein} is that contributions involving $M'$ have cancelled in intermediate steps; we anticipate that this will not be true at one loop.

It is useful to notice that, in contrast to the tree-level contribution, the amplitude fragment~\eqref{quantrein} is purely imaginary (or $i\tilde{\mathcal{A}}(b)$ is real). 
Hence, each fragment corresponds to real and imaginary parts of the total tree-level amplitude
\begin{align}
\label{eq:RealImPartAmp}
\text{Re}[\tilde{\mathcal{A}}_0(b)] = \tilde{\mathcal{A}}_0^{(0)}(b) + \cdots \,,\qquad
i\,\text{Im}[\tilde{\mathcal{A}}_0(b)] = \tilde{\mathcal{A}}_0^{(1)}(b) + \cdots \,.
\end{align}
We hope it is clear to the reader that these are the real and imaginary parts of the impact-parameter space amplitude rather than the usual momentum-space ones; we have checked that this even/odd pattern persists at higher orders.
The  equation above is one of the consequences of scattering in a time-varying background, because in static Schwarzschild the tree amplitude is purely real.
In the Schwarzschild case, the tree amplitude is real but it contains a subdominant quantum (contact) term. Here we have an imaginary subleading term.
It arises in the computation due to both the $\Theta(\sfk\cdot x)$ step function and to the different boundary conditions: ${b_{\rm Schw.}^\mu = (0,\bm{b})}$ instead of ${b^\mu_{\rm Vaidya} = (v,\bm{0})}$. 

In the next section, we will see the importance of this  correction for the NLO one-loop phase.

\subsection{Subleading part of the one-loop amplitude} 

Let us now move to the one loop case. The amplitude in momentum space is
\begin{align}\label{eq:1loopamp}
i\cA_{1}(p-q\to p) = -i
\!\!\int\!\hat{\dd}^4\ell\, \frac{\sfh_{\mu\nu}(\ell)\,\sfh_{\rho\sigma}
(q-\ell)}{(p-q+\ell)^2+i\epsilon}
\left[
 (p-q)^\mu (p-q+\ell)^\nu
\right]
\left[p^\rho (p-q+\ell)^\sigma 
\right],
\end{align}
which corresponds to the diagram below  

\[\label{eq:OneLoopDiagram}
\begin{tikzpicture}[x=0.75pt,y=0.75pt,yscale=-1,xscale=1]
  
\tikzset {_t5hf34dje/.code = {\pgfsetadditionalshadetransform{ \pgftransformshift{\pgfpoint{0 bp } { 0 bp }  }  \pgftransformrotate{0 }  \pgftransformscale{2 }  }}}
\pgfdeclarehorizontalshading{_t6j2i7lvl}{150bp}{rgb(0bp)=(1,1,1);
rgb(42.606724330357146bp)=(1,1,1);
rgb(45.89285714285714bp)=(0.82,0.89,0.97);
rgb(57.14285714285714bp)=(0.29,0.56,0.89);
rgb(57.25bp)=(0.29,0.56,0.89);
rgb(62.5bp)=(0.29,0.56,0.89);
rgb(100bp)=(0.29,0.56,0.89)}

  
\tikzset {_9bzzwff3a/.code = {\pgfsetadditionalshadetransform{ \pgftransformshift{\pgfpoint{0 bp } { 0 bp }  }  \pgftransformrotate{0 }  \pgftransformscale{2 }  }}}
\pgfdeclarehorizontalshading{_70ex34ubd}{150bp}{rgb(0bp)=(1,1,1);
rgb(42.428152901785715bp)=(1,1,1);
rgb(47.67857142857143bp)=(0.82,0.89,0.97);
rgb(52.23214285714286bp)=(0.29,0.56,0.89);
rgb(57.5bp)=(0.29,0.47,0.89);
rgb(62.42815290178571bp)=(0.29,0.47,0.89);
rgb(100bp)=(0.29,0.47,0.89)}

  
\tikzset {_bjh1003yi/.code = {\pgfsetadditionalshadetransform{ \pgftransformshift{\pgfpoint{0 bp } { 0 bp }  }  \pgftransformrotate{0 }  \pgftransformscale{2 }  }}}
\pgfdeclarehorizontalshading{_1y3sbix1p}{150bp}{rgb(0bp)=(1,1,1);
rgb(42.606724330357146bp)=(1,1,1);
rgb(45.89285714285714bp)=(0.82,0.89,0.97);
rgb(57.14285714285714bp)=(0.29,0.56,0.89);
rgb(57.25bp)=(0.29,0.56,0.89);
rgb(62.5bp)=(0.29,0.56,0.89);
rgb(100bp)=(0.29,0.56,0.89)}

  
\tikzset {_ab4uab8zc/.code = {\pgfsetadditionalshadetransform{ \pgftransformshift{\pgfpoint{0 bp } { 0 bp }  }  \pgftransformrotate{0 }  \pgftransformscale{2 }  }}}
\pgfdeclarehorizontalshading{_toll77psf}{150bp}{rgb(0bp)=(1,1,1);
rgb(42.428152901785715bp)=(1,1,1);
rgb(47.67857142857143bp)=(0.82,0.89,0.97);
rgb(52.23214285714286bp)=(0.29,0.56,0.89);
rgb(57.5bp)=(0.29,0.47,0.89);
rgb(62.42815290178571bp)=(0.29,0.47,0.89);
rgb(100bp)=(0.29,0.47,0.89)}
\tikzset{every picture/.style={line width=0.75pt}} 

\draw    (110,17145) -- (156.02,17152.46) ;
\draw [shift={(135.58,17149.15)}, rotate = 189.21] [fill={rgb, 255:red, 0; green, 0; blue, 0 }  ][line width=0.08]  [draw opacity=0] (5.36,-2.57) -- (0,0) -- (5.36,2.57) -- cycle    ;
\draw    (229.82,17152.69) -- (277,17144) ;
\draw [shift={(255.97,17147.88)}, rotate = 169.56] [fill={rgb, 255:red, 0; green, 0; blue, 0 }  ][line width=0.08]  [draw opacity=0] (5.36,-2.57) -- (0,0) -- (5.36,2.57) -- cycle    ;
\draw    (156.02,17152.46) -- (229.82,17152.69) ;
\draw [shift={(195.52,17152.58)}, rotate = 180.18] [fill={rgb, 255:red, 0; green, 0; blue, 0 }  ][line width=0.08]  [draw opacity=0] (5.36,-2.57) -- (0,0) -- (5.36,2.57) -- cycle    ;
\draw    (164.6,17170.62) -- (164.6,17183.62) ;
\draw [shift={(164.6,17167.62)}, rotate = 90] [fill={rgb, 255:red, 0; green, 0; blue, 0 }  ][line width=0.08]  [draw opacity=0] (5.36,-2.57) -- (0,0) -- (5.36,2.57) -- cycle    ;
\draw    (241.4,17171.62) -- (241.4,17184.62) ;
\draw [shift={(241.4,17168.62)}, rotate = 90] [fill={rgb, 255:red, 0; green, 0; blue, 0 }  ][line width=0.08]  [draw opacity=0] (5.36,-2.57) -- (0,0) -- (5.36,2.57) -- cycle    ;
\draw    (232.32,17152.69) .. controls (233.99,17154.35) and (234,17156.02) .. (232.34,17157.69) .. controls (230.68,17159.36) and (230.68,17161.02) .. (232.35,17162.69) .. controls (234.02,17164.35) and (234.03,17166.02) .. (232.37,17167.69) .. controls (230.71,17169.36) and (230.71,17171.02) .. (232.38,17172.69) .. controls (234.05,17174.35) and (234.06,17176.02) .. (232.4,17177.69) .. controls (230.74,17179.36) and (230.74,17181.02) .. (232.41,17182.69) .. controls (234.08,17184.36) and (234.08,17186.02) .. (232.42,17187.69) .. controls (230.76,17189.36) and (230.77,17191.03) .. (232.44,17192.69) .. controls (234.11,17194.36) and (234.11,17196.02) .. (232.45,17197.69) .. controls (230.79,17199.36) and (230.8,17201.03) .. (232.47,17202.69) -- (232.48,17205.91) -- (232.48,17205.91)(229.32,17152.7) .. controls (230.99,17154.36) and (231,17156.03) .. (229.34,17157.7) .. controls (227.68,17159.37) and (227.68,17161.03) .. (229.35,17162.7) .. controls (231.02,17164.36) and (231.03,17166.03) .. (229.37,17167.7) .. controls (227.71,17169.37) and (227.71,17171.03) .. (229.38,17172.7) .. controls (231.05,17174.36) and (231.06,17176.03) .. (229.4,17177.7) .. controls (227.74,17179.37) and (227.74,17181.03) .. (229.41,17182.7) .. controls (231.08,17184.37) and (231.08,17186.03) .. (229.42,17187.7) .. controls (227.76,17189.37) and (227.77,17191.04) .. (229.44,17192.7) .. controls (231.11,17194.37) and (231.11,17196.03) .. (229.45,17197.7) .. controls (227.79,17199.37) and (227.8,17201.04) .. (229.47,17202.7) -- (229.48,17205.92) -- (229.48,17205.92) ;
\path  [shading=_t6j2i7lvl,_t5hf34dje] (230.58,17201.71) .. controls (225.93,17201.71) and (222.16,17205.23) .. (222.16,17209.58) .. controls (222.16,17213.93) and (225.93,17217.46) .. (230.58,17217.46) .. controls (235.23,17217.46) and (239,17213.93) .. (239,17209.58) .. controls (239,17205.23) and (235.23,17201.71) .. (230.58,17201.71) -- cycle ; 
 \draw   (230.58,17201.71) .. controls (225.93,17201.71) and (222.16,17205.23) .. (222.16,17209.58) .. controls (222.16,17213.93) and (225.93,17217.46) .. (230.58,17217.46) .. controls (235.23,17217.46) and (239,17213.93) .. (239,17209.58) .. controls (239,17205.23) and (235.23,17201.71) .. (230.58,17201.71) -- cycle ; 

\path  [shading=_70ex34ubd,_9bzzwff3a] (230.42,17201.25) .. controls (235.04,17201.27) and (238.77,17204.93) .. (238.74,17209.42) .. controls (238.72,17213.9) and (234.95,17217.52) .. (230.32,17217.5) .. controls (225.7,17217.48) and (221.97,17213.82) .. (222,17209.33) .. controls (222.03,17204.84) and (225.8,17201.22) .. (230.42,17201.25) -- cycle ; 
 \draw   (230.42,17201.25) .. controls (235.04,17201.27) and (238.77,17204.93) .. (238.74,17209.42) .. controls (238.72,17213.9) and (234.95,17217.52) .. (230.32,17217.5) .. controls (225.7,17217.48) and (221.97,17213.82) .. (222,17209.33) .. controls (222.03,17204.84) and (225.8,17201.22) .. (230.42,17201.25) -- cycle ; 

\draw    (157.52,17152.45) .. controls (159.19,17154.12) and (159.19,17155.78) .. (157.53,17157.45) .. controls (155.87,17159.12) and (155.88,17160.79) .. (157.55,17162.45) .. controls (159.22,17164.12) and (159.22,17165.78) .. (157.56,17167.45) .. controls (155.9,17169.12) and (155.91,17170.79) .. (157.58,17172.45) .. controls (159.25,17174.12) and (159.25,17175.78) .. (157.59,17177.45) .. controls (155.93,17179.12) and (155.93,17180.78) .. (157.6,17182.45) .. controls (159.27,17184.11) and (159.28,17185.78) .. (157.62,17187.45) .. controls (155.96,17189.12) and (155.96,17190.78) .. (157.63,17192.45) .. controls (159.3,17194.11) and (159.31,17195.78) .. (157.65,17197.45) .. controls (155.99,17199.12) and (155.99,17200.78) .. (157.66,17202.45) -- (157.67,17205.67) -- (157.67,17205.67)(154.52,17152.46) .. controls (156.19,17154.13) and (156.19,17155.79) .. (154.53,17157.46) .. controls (152.87,17159.13) and (152.88,17160.8) .. (154.55,17162.46) .. controls (156.22,17164.13) and (156.22,17165.79) .. (154.56,17167.46) .. controls (152.9,17169.13) and (152.91,17170.8) .. (154.58,17172.46) .. controls (156.25,17174.13) and (156.25,17175.79) .. (154.59,17177.46) .. controls (152.93,17179.13) and (152.93,17180.79) .. (154.6,17182.46) .. controls (156.27,17184.12) and (156.28,17185.79) .. (154.62,17187.46) .. controls (152.96,17189.13) and (152.96,17190.79) .. (154.63,17192.46) .. controls (156.3,17194.12) and (156.31,17195.79) .. (154.65,17197.46) .. controls (152.99,17199.13) and (152.99,17200.79) .. (154.66,17202.46) -- (154.67,17205.68) -- (154.67,17205.68) ;
\path  [shading=_1y3sbix1p,_bjh1003yi] (155.58,17201.71) .. controls (150.93,17201.71) and (147.16,17205.23) .. (147.16,17209.58) .. controls (147.16,17213.93) and (150.93,17217.46) .. (155.58,17217.46) .. controls (160.23,17217.46) and (164,17213.93) .. (164,17209.58) .. controls (164,17205.23) and (160.23,17201.71) .. (155.58,17201.71) -- cycle ; 
 \draw   (155.58,17201.71) .. controls (150.93,17201.71) and (147.16,17205.23) .. (147.16,17209.58) .. controls (147.16,17213.93) and (150.93,17217.46) .. (155.58,17217.46) .. controls (160.23,17217.46) and (164,17213.93) .. (164,17209.58) .. controls (164,17205.23) and (160.23,17201.71) .. (155.58,17201.71) -- cycle ; 

\path  [shading=_toll77psf,_ab4uab8zc] (155.42,17201.25) .. controls (160.04,17201.27) and (163.77,17204.93) .. (163.74,17209.42) .. controls (163.72,17213.9) and (159.95,17217.52) .. (155.32,17217.5) .. controls (150.7,17217.48) and (146.97,17213.82) .. (147,17209.33) .. controls (147.03,17204.84) and (150.8,17201.22) .. (155.42,17201.25) -- cycle ; 
 \draw   (155.42,17201.25) .. controls (160.04,17201.27) and (163.77,17204.93) .. (163.74,17209.42) .. controls (163.72,17213.9) and (159.95,17217.52) .. (155.32,17217.5) .. controls (150.7,17217.48) and (146.97,17213.82) .. (147,17209.33) .. controls (147.03,17204.84) and (150.8,17201.22) .. (155.42,17201.25) -- cycle ; 

\draw (67,17135) node [anchor=north west][inner sep=0.75pt]    {$p-q$};
\draw (282,17136) node [anchor=north west][inner sep=0.75pt]    {$p$};
\draw (172.6,17167.62) node [anchor=north west][inner sep=0.75pt]    {$\ell $};
\draw (160,17129) node [anchor=north west][inner sep=0.75pt]    {$p-q+\ell $};
\draw (247.6,17167.62) node [anchor=north west][inner sep=0.75pt]    {$q-\ell $};
\draw (303,17167.5) node [anchor=north west][inner sep=0.75pt]    {$=i\mathcal{A}_1(p-q\to p).$};

\end{tikzpicture}
\]

As we are computing the eikonal in position space, subleading contributions to the phase also arise from the expansion of the $e^{i\lambda q^2}$ factors in the on-shell delta functions as in equation~\eqref{qsqq}.

Moreover, it will also be convenient to reorganize one-loop expressions by relabelling the loop momenta as $\ell=\ell_1$ and ${q-\ell=\ell_2}$; then the expressions and propagators take a more symmetric form. 
With this in mind we expand the one-loop amplitude following the hierarchy of momenta dictated by equation~\eqref{eq:geolocks}. 
Practically, this means Laurent-expanding the propagators and numerators in equation~\eqref{eq:1loopamp} for small $ \ell_{1,2}$. 
Furthermore, we deal with loop variables in propagators via Schwinger parameters as follows\footnote{As the energy $p^0$ is large and positive in the geometric optics approximation, we may neglect here the distinction between retarded and Feynman propagators. As we will not consider crossing this contribution, we need not be concerned with any associated subtleties.}
\[\label{eq:schtrick}
\frac{ i}{2p\cdot \ell_i+ i\epsilon}=\int_{-\infty}^\infty \dd\lambda_i \Theta(\lambda_i)e^{ i\lambda_i(2p\cdot \ell_i + i\epsilon)},\,\,\,\, \frac{-1}{(2p\cdot \ell_i+ i\epsilon)^2}=\int_{-\infty}^\infty \dd\lambda_i\Theta(\lambda_i)\lambda_i e^{i\lambda_i(2p\cdot \ell_i + i\epsilon)},
\]
with $i={1,2}.$
The amplitude in position space can then be written as
\begin{align}
\label{eq:OneLoopAmp_Symmetric}
i\tilde{\mathcal{A}}_1(b) 
    =\frac{i^2}{2} \int 
    [\dd\lambda_1\, \hat{\dd}^4\ell_1e^{-i(b+2\lambda_1 p )\cdot \ell_1}] 
    [\dd\lambda_2\, \hat{\dd}^4\ell_2e^{-i(b+2\lambda_2 p )\cdot \ell_2}] 
    \,{\cal J}(\ell_1,\lambda_1,\ell_2,\lambda_2),
\end{align}
for some function ${\cal J}(\ell_1,\lambda_1,\ell_2,\lambda_2)$ that will be presented below.
 
To better organise our results, in equation \eqref{eq:OneLoopAmp_Symmetric} we divide contributions according to their scaling in the geometric-optics limit
\begin{align}
    {\cal J}= {\cal J}^{(-1)} + {\cal J}^{(0)} + {\cal J}^{(1)} +\cdots\,.
\end{align}
Here, ${\cal J}^{(-1)}$ is the dominant iteration term\footnote{This is the analogue of the ``superclassical'' term in the usual classical scattering problem.} discussed previously in section~\ref{sec:Tree_level}, namely
\begin{align}
{\cal J}^{(-1)} (\ell_1, \ell_2) \equiv  (p\cdot   \sfh(\ell_1)\cdot p) \,(p\cdot   \sfh(\ell_2)\cdot p).
\end{align}
Since it is immediately clear that this term yields a perfect square in position space, we move on to the subleading contribution
\[
{\cal J}^{(0)}  \equiv  {\cal J}^{(0)}_{\rma} +{\cal J}^{(0)}_{\rmb} \,.
\]
We divide this contribution into two parts, according to whether the Heaviside functions can be combined into unity or not using the identity $\Theta(\lambda) + \Theta(-\lambda) =1$.
The reasons for using this splitting will also be clearer below.  
Then, we find
\begin{subequations}
\label{eq:OneLoop_JaJb}
\begin{align}
    {\cal J}_\rma^{(0)} 
    &\equiv {\cal J}^{(-1)}  (\ell_1, \ell_2)  \bigg[i\lambda_1\ell_1^2 + i\lambda_2\ell_2^2\bigg]
    {-}\,
   \sfh_{\mu\nu}(\ell_2)\sfh_{\rho \sigma}(\ell_1)\bigg[p^\mu p^\nu p^\rho \ell_1^\sigma + p^\mu \ell_2^\nu p^\rho  p^\sigma\bigg],\\
    {\cal J}_\rmb^{(0)}  &\equiv
    2\,\Theta(\lambda_1-\lambda_2)\,(p\cdot \sfh( \ell_1) \cdot p)  \,
    \sfh_{\mu\nu}(\ell_2)\bigg[
     i\lambda_2(\ell_1\cdot \ell_2) p^\mu p^\nu -\ell_1^\mu p^\nu
    \bigg] + (1{\leftrightarrow} 2),
\end{align}
\end{subequations}
where $(1\leftrightarrow 2)$ accounts for swapping $\lambda_i$ and $\ell_i$'s for $i=1,2$.
As in the tree-level case, we now use IBP to trade $\ell$'s for derivatives acting on the metric through equations \eqref{FTh} and \eqref{eq:Derivative_h}. Then, we trivially evaluate the loop integrals in equation~\eqref{eq:OneLoopAmp_Symmetric}, which are at this point pure Fourier exponents: they simply impose the constraints that all spacetime dependent quantities are evaluated at 
\[
x^\mu_{1,2}=b^\mu(v)+2\lambda_{1,2}\, p^\mu.
\]
Finally, one is left with two one-dimensional $\lambda_{1,2}$ integrals. After evaluating these we find the surprisingly simple result for the one-loop subleading fragment\footnote{We reproduced equation \eqref{eq:OneLoopAmp} with an automated code as well as comparing with the results of \cite{Menezes:2022tcs}, where a similar one-loop calculation was tackled using Kerr-Schild coordinates for the background. See also \cite{Bianchi:2023lrg} for more on the KS gauge and perturbation theory.}
\begin{align}
\label{eq:OneLoopAmp}
    i\tilde{\mathcal{A}}^{(0)}_1(b)   
    &=  -i16(GM)^2\frac{u\cdot p}{u\cdot b}\bigg[1+\log\left(\frac{4u\cdot  p \, \lambda_{\infty}}{-u\cdot b}\right)
\bigg],
\end{align}
which diverges logarithmically for large times as $\lambda_{\infty} \rightarrow \infty$. 
This is  important. 
The divergence is now a physical one because, in contrast with the tree-level divergence in equation~\eqref{eq:intlamtree}, it depends on $b^\mu$ and would in principle affect any calculable observable.  
We seem to have an issue here. 
However, we know that  the Hawking amplitude in Eq. \eqref{eq:AHawking} is well-defined and finite at this order, so a cancellation must occur.
Indeed, in the following subsection we will show that the divergent term above is neatly removed when tree-level iterations are properly accounted for.

\subsection{The eikonal versus the amplitude} 
Let us pause for a moment the direct evaluation of the NLO phase  to remind ourselves how  $\tilde{\mathcal{A}}(b)$ and the eikonal are related. In~\cite{DiVecchia:2023frv, Cristofoli:2021jas}, the standard eikonal exponentiation is given as
\begin{align}
\label{eq:EikonalRelations}	
    e^{i\chi(b)/\hbar} \left(1+ i \hbar\Delta(b)\right) = 1+ i \tilde{\mathcal{A}}(b).
\end{align}
This relationship between the amplitude and the phase  involves a generic function $i\Delta(b)\equiv i \Delta(p,b)$ which entails corrections that do not exponentiate. In  a post-Minkowskian expansion these corrections  simply represent quantum terms, as a consequence $i\Delta(b)$ is usually referred to as the ``quantum remainder" of a given theory. 
This nomenclature is not completely accurate in our case: ours is a ray optics expansion, and so we will simply refer to  $i\Delta(b)$  as the ``remainder". 

Next, it is useful to expand each term in powers of the effective coupling $(GM)$ as follows
\begin{align}
i\chi(b) = (GM) i\chi_0(b) + (GM)^2 i\chi_1(b) +\cdots  = \sum_{n=0}^\infty (GM)^{n+1}i\chi_n(b),
\end{align}
and similarly for $i\Delta(b)$ and $i\tilde{\mathcal{A}}(b)$. In the post-Minkowskian literature,  quantum terms are usually discarded given the main interest in  classical observables. What this will show is that the $\hbar$ counting is such that lower iterations of $\Delta$ will enter the calculation of a given eikonal fragment. For instance, the $\Delta_2$ piece can be relevant at two loops through iterations that compute $\chi_2$, on the other hand  $\Delta_0$ and $\Delta_1$ vanish for   Schwarzschild scattering. This pattern was first discussed in~\cite{DiVecchia:2021bdo}.
The vanishing of the Schwarzschild remainder -- in  our framework --   can be interpreted in terms of the lack of a scale in the $\dd\lambda$ integral at play. Instead, considering the Vaidya spacetime naturally introduces the equivalent of a  position space ``cut-off"  -- defined by the step function $\Theta(\sfk\cdot x)$ -- which leaves out a finite contribution already at leading order.

With this in mind, we move back to formula \ref{eq:EikonalRelations} and expand both sides. We have
\begin{align}
1+ \left(i\chi_0 + i\Delta_0\right) + \frac{1}{2!}\left(i\chi_0\right)^2 & + \left(i\chi_1+ i\Delta_0i\chi_0+i\Delta_1 \right) + \cdots  \\ &=1+\left(i\tilde{\mathcal{A}}^{(0)}_0+i\tilde{\mathcal{A}}^{(1)}_0 \right)+ 
    \left(
    i\tilde{\mathcal{A}}^{(-1)}_1 + i\tilde{\mathcal{A}}^{(0)}_1 + i\tilde{\mathcal{A}}^{(1)}_1
    \right)
    + \cdots\notag
\end{align}
One can then immediately equate terms on both sides of the equal sign  order by order. For instance, the leading order eikonal and amplitude are straightforwardly related and the leading one loop piece is a tree-level  (leading) iteration 
\begin{equation}
i\chi_0 = i\tilde{\mathcal{A}}^{(0)}_0,\,\,\,\,\,\,\,\,\frac{1}{2!}(\chi_0)^2 = i\tilde{\mathcal{A}}^{(-1)}_1,
\end{equation}
which is consistent with what we have seen before. 
However, on a Vaidya background there is a new  ingredient:   the non-vanishing  remainder of the tree level amplitude. This implies that, at one-loop we have
\begin{align}\label{relation}
i\Delta_0 = i\tilde{\mathcal{A}}^{(1)}_0 \neq 0
\qquad\Rightarrow\qquad
i\chi_1+ i\Delta_0i\chi_0 = i\tilde{\mathcal{A}}^{(0)}_1,
\end{align}
which relates then -- in a non-trivial way  -- the one loop amplitude (which we found to diverge by itself) and the tree-level one. Let us see what   the implications are next.

\subsection{The finite NLO eikonal phase}

As the reader may have already noticed, this iteration exactly cancels the log divergent term found in Eq.~\eqref{eq:OneLoopAmp}.
Plugging the position space fragments computed earlier in equations   \eqref{frag0},~\eqref{quantrein} and~\eqref{eq:OneLoopAmp} into the relation \eqref{relation}, we finally obtain the next-to-leading order eikonal phase 
\begin{align}
\label{eq:chi1}
i\chi_1 
&= i\tilde{\mathcal{A}}^{(0)}_1 - i\tilde{\mathcal{A}}^{(1)}_0 i\tilde{\mathcal{A}}^{(0)}_0\nonumber\\
&=-i16(GM)^2\frac{u\cdot p}{u\cdot b}\bigg[1+\log\left(\frac{4u\cdot  p\,  \lambda_{\infty}}{-u\cdot b} \right)\!\!\bigg]
{-}
i4GM u\cdot p\log\left(\frac{4u\cdot  p\,  \lambda_{\infty}}{-u\cdot b}\right)\! \left[\frac{-4GM}{u\cdot b}\right] \nonumber\\
&=-i16(GM)^2\frac{u\cdot p}{u\cdot b} ,
\end{align}
leaving a finite NLO eikonal, as it should. 
Crucial to this cancellation is the fact that $\tilde{\mathcal{A}}^{(1)}_0$ is imaginary (see equation~\ref{eq:RealImPartAmp}). 
This implies that the product $i\tilde{\mathcal{A}}^{(1)}_0 i\tilde{\mathcal{A}}^{(0)}_0$ is also imaginary.

The cancellation of this logarithmic divergence is actually more general than we have shown so far.
This can be better appreciated if the iteration term is rearranged in the form of a more familiar loop contribution.
In fact, this is at the integrand level 
\begin{align}\label{divlog}
i\tilde{\mathcal{A}}^{(1)}_0 i\tilde{\mathcal{A}}^{(0)}_0 
= 
    i^2 \int 
    [\dd\lambda_1\, \hat{\dd}^4\ell_1&e^{-i(b+2\lambda_1 p )\cdot \ell_1}] 
    [\dd\lambda_2\,\hat{\dd}^4\ell_2e^{-i(b+2\lambda_2 p )\cdot \ell_2}] \\
    &\times p\cdot \sfh( \ell_1)\cdot p \,
   \sfh_{\mu\nu}(\ell_2)\, \left(i\lambda_2\, \ell_2^2 \,p^\mu p^\nu-p^\mu \ell_2^{\nu}\right),\nn
\end{align}
where we labelled $\tilde{\mathcal{A}}^{(0)}$ ($\tilde{\mathcal{A}}^{(1)}$) term with $\ell_1 \, (\ell_2)$. 
One can verify that equation~\eqref{divlog} is purely divergent. 
Now, if we symmetrize over $\ell_1\leftrightarrow \ell_2$ and average over the two expressions, we see immediately that we get exactly the integrand defining $\mathcal{J}^{(0)}_\rma$. 
Thus, the cancellation is already realised at the integrand level
\begin{align}
   \mathcal{J}^{(0)} - \mathcal{J}^{\rm iter} = 
    \mathcal{J}^{(0)}_\rma + 
    \mathcal{J}^{(0)}_\rmb - \mathcal{J}^{\rm iter} 
    = \mathcal{J}^{(0)}_\rmb,
\end{align}
where $\mathcal{J}^{\rm iter}$ is the integrand defined by the iterations in Eq. \eqref{divlog}. 
As we can see, this subtraction leaves only the purely finite term $\mathcal{J}^{(0)}_\rmb$.  
This rearrangement is the reason why we split $\mathcal{J}^{(0)}$ in equation \eqref{eq:OneLoopAmp} to begin with. 
This early cancellation also allows us to show another important property: the final result is finite independently of the particular mass function $M(v)$. 
In other words, it is independent of how the black hole was formed. 
Simultaneously, we can say more about $\mathcal{J}^{(0)}_\rmb$. 
For this term, it happens that after the loop integration we are left with terms either proportional to $M(v_1)M(v_2)$ or to $M(v_{1,2})M'(v_{2,1})$.
The last ones arise through derivatives acting on the metric as explained in equation~\eqref{eq:Derivative_h}. 
It turns out that the general (that is, $M(v)$ dependent) subleading eikonal reads 
\begin{align}
\label{chi1afterints}
    \chi_1 &=
     8G^2(u\cdot p)
    \int \dd\lambda_1\,\dd\lambda_2\,
    \Theta(\lambda_1)\Theta(\lambda_2)
    \\&
    \qquad\qquad\times \bigg(\frac{1}{\lambda_1}\Theta(\lambda_1-\lambda_2) + \frac{1}{\lambda_2}\Theta(\lambda_2-\lambda_1)\bigg)
    \bigg(\frac{M(v_1)M'(v_2)}{\lambda_1}+
    \frac{M(v_2)M'(v_1)}{\lambda_2}
    \bigg),\nn\\&
    =
     16G^2(u\cdot p)
    \int \dd\lambda_1\,\dd\lambda_2\,
\Theta(\lambda_1)\Theta(\lambda_2)\bigg(\frac{\Theta(\lambda_1-\lambda_2)}{\lambda_1^2}{M(v_1)M'(v_2)} \nn\\& 
\hspace{4.5cm}+\frac{\Theta(\lambda_1-\lambda_2)}{\lambda_1^2}\frac{ M(v_1)M(v_2)}{4u\cdot p\,\lambda_2}
    -\frac{M(v_1)M(v_2)}{4u\cdot p\,\lambda_2^2}\delta(\lambda_1-\lambda_2)\bigg), \nn
\end{align}
where the second expression was obtained by rewriting the crossed terms scaling as $\sim 1/(\lambda_1\lambda_2)$ via integration by parts. 
In fact, note that now the NLO eikonal scales at worst like $\mathcal{O}( M(v_1)\log\lambda_1/\lambda^2_1)$ for large $\lambda_1$ values. 
This shows that equation~\eqref{chi1afterints} is convergent provided that the mass function is bounded by some positive constant ${|M(v)|\leq M}$ and that its derivative $M'(v)$ is sharply localised around some $v=\bar v$. 
Said differently, we expect that under the assumption that the collapse occurs relatively fast compared to the timescales of $\lambda$, the crossed terms  vanish. 
Finally, note that when ${M(v)=M \Theta(v)}$ equation~\eqref{chi1afterints} returns the result we found previously in equation~\eqref{eq:chi1}. 
However, we stress again that this expression is more general and applies to any (reasonable) collapse model.

Let us now go back to the equivalence between the amplitude and the classical derivation. Above, we have  obtained $i\chi_1$ perturbatively, and this  should  account for the   expansion of the horizon term in the $\log(v-v_0)$ of section~\ref{sec:Review}. 
This is indeed true. 
One can see this by resumming~\eqref{eq:chi1} with the LO eikonalization of equation~\eqref{eq:ResumedAmp} at the consistent order. 
Conversely, we can compare to the expanded amplitude in \eqref{eq:AHawking} with the expansion of the amplitude's eikonal
\begin{align}
\bra{p}S_{\rm L+NL}\ket{\psi}
&=\int \dd v \,\varphi(v) e^{i p\cdot b(v)} \left( 1+ i\chi_0+\frac{1}{2!}(i\chi_0)^2 +  i\chi_1 + \cdots \right)\\
&=\frac{2\pi}{E_0}
\int_{-\infty}^{0} \dd v 
\left( 1+ i\chi_0+\frac{1}{2!}(i\chi_0)^2 +  i\chi_1 + i(E-E_0)v +\cdots \right)\nn
\end{align}
where in the second step we have chosen, as in section~\ref{sec:Tree_level},
$\varphi(v) =  2 \pi /E_0 \,e^{-iE_0v}$ and ${b(v)\cdot p = Ev}$. The equivalence between the two expressions at $\mathcal{O}((GM)^2)$ is immediately established by choosing a frame where $u\cdot p =E $ and $u\cdot b=v$ inside the formulas found for  $\chi_0$ and $\chi_1$.

Finally, we end this section by noting that ultimately loop corrections do not change the form of the distribution discussed in section \ref{spectrum}, which only used tree-level data.
This is immediate if we return to the full all-order Hawking amplitude, equation~\eqref{eq:Ahawkfull} which we reproduce here for convenience:
\begin{equation}
\braket{p|S|\psi} = \int \dd v \,\varphi(v) e^{i p\cdot b(v)} e^{-4iGME\log(v_0-v)}.
\end{equation}
As in section \ref{spectrum} we see that the $v$ integral is only well-defined over a restricted domain, dictated by the  position of the branch cut. 
This time it makes sense to integrate for all  values of $v<v_0=-4GM$.
Now consider shifting the integration variable to $v'=v-v_0$. 
Then, one can immediately establish that
\[
   \mathcal{A} &=
  e^{-4iGM(E-E_0)}  \mathcal{A}^{\rm tree} \,.
\]
Thus, the difference between the loop corrected and the tree-level  amplitude is nothing but a pure phase, but  we know this is unimportant since the spectrum  only depends on $|\mathcal{A}|$.
We will comment further on this point in the conclusions.

\section{Discussion and Outlook}
\label{sec:Conclusions}

In this paper, we reinterpreted Hawking's classical computation~\cite{Hawking:1974rv,Hawking:1975vcx} using the language of present-day scattering amplitudes.
However, we hope that our work will lead to future developments in our understanding of black hole physics because, as we will discuss shortly, there is a clear route to reformulating this process in flat space quantum field theory throughout.

First let us briefly summarise  this article. 
The key step was a resummation of Feynman diagrams representing the scattering of a massless scalar in a collapse background. 
Indeed, resummation is an important aspect of the amplitudes/gravitational waves programmes so the structure of the resummation was very familiar.
One interesting difference relative to (for example) eikonal exponentiation was that our expansion was in the geometric optics limit, resulting in what we call the Hawking amplitude.
The exponentiation in the geometric optics limit is the diagrammatic version of the original ray tracing argument. 

We demonstrated that our resummed generalised amplitude is the Bogoliubov $A(p,k)$ coefficient in section \ref{sec:Bogoliubov}. 
Remarkably, the importance of generalised amplitudes was highlighted very recently by~\cite{Caron-Huot:2023vxl,Caron-Huot:2023ikn}.
We connected the well-known analytic continuation relating the Bogoliubov $A$ and $B$ coefficients to crossing in the usual sense of quantum field theory.

In section~\ref{sec:OneLoop} we further confirmed our scattering methods by computing the subleading correction to the phase. 
This subleading correction is interesting because it is directly sensitive to the finite radius of the black hole, so in this sense it is a finite-size correction.
The calculation was non-trivial, and we leaned heavily on our knowledge of the systematics of the eikonal expansion~\cite{DiVecchia:2023frv}.
Naively, our subleading eikonal is divergent: to resolve this impasse we had to carefully subtract a specific part of the tree-level amplitude from the NLO phase.
This kind of subtraction is familiar in (for example) Schwarzschild scattering, but there it only arises at higher loop orders.

Now let us discuss our vision for future directions following this work.
Some directions are immediately obvious.
For example, it is important to investigate different black hole backgrounds such as Reissner-Nordstr\"om and Kerr.
In the spirit of scattering amplitudes, it is also interesting to wonder about the double copy.
Is there an electromagnetic system which is related to Hawking's process by the double copy?
Our work makes it clear that mathematically this should work; however the physical interpretation of the single copy of the Vaidya background seems very unclear.

From our perspective, a very intriguing aspect of the ray-tracing relation, equation~\eqref{eq:rayTracingRelation}, is that all its dependence on $v_0$ can be removed by a translation: defining $v' = v- v_0$. 
Then the exact Hawking amplitude from equation~\eqref{eq:Ahawkfull} becomes
\[
\mathcal{A}_H 
&= 
\frac{2 \pi}{E} \int_{-\infty}^{0} \dd v' \, 
\exp \left[-4 i GM \, E \log \big( - v' / \mu \big) + i (E - E_0) (v' - 4 GM)\right] \,.
\]
Observe that the phase is now exactly linear in $GM$; all finite size effects are removed by the translation.
This is why the exponentiation of the tree approximation to the phase yields the exact thermal distribution.
The phenomenon that the tree approximation to some quantity differs from an exact computation of the same quantity by a translation occurs elsewhere.
A prominent example is the Schwarzschild metric in Kerr-Schild form (which is a linearised gauge transformation of the linearised solution).
Similarly, the important Veneziano-Vilkovisky large gauge transformation corresponds to exponentiating three-point amplitudes \cite{Elkhidir:2024izo}.
Are these a coincidence or is there some underlying structure at work? 
The notion that general relativity is not-so-nonlinear~\cite{Harte:2014ooa} could be relevant.

Finally, as we suggested above, our work lends itself to an interpretation in terms of flat space scattering amplitudes.
We worked in a Vaidya background.
This involves, on $\scri^-$, some infalling lightlike radiation.
In quantum field theory we can capture this radiation by introducing a new massless scalar quantum field $\varphi(x)$ which is in a coherent state $\ket{\alpha}$ in the far past.
We can choose $\ket\alpha$ to be a spherically symmetric field of infalling radiation which is localised at $v=0$ on $\scri^-$.
Once the number of quanta in the coherent state is large enough, this coherent state can be understood as a classical field.
If the energy density is large enough when the field falls towards the origin, a black hole will inevitably form.
This is a concrete recipe for embedding the Vaidya background in the language of flat space amplitudes.

It is helpful to think of the coherent state in terms of a displacement operator $\ket \alpha = \mathbb{C}_\alpha \ket{0}$ where 
\[
\mathbb{C}_\alpha = \exp \left[ \int \dd \Phi(k) \alpha(k) \, (a_\varphi^\dagger(k) - a_\varphi(k) )\right] \,.
\]
We chose a real parameter function $\alpha(k)$ (see reference~\cite{Cristofoli:2021vyo} for a recent discussion of coherent states in the context of amplitudes and gravitational waves).
Here $a_\varphi(k)$ is a ladder operator for the $\varphi(x)$ field. 
Displacement operators have the property that the $S$-matrix in the coherent background can be written as
\[
S[\alpha] = \mathbb{C}_\alpha^\dagger S[0] \mathbb{C}_\alpha \,.
\]
The notation here is that $S[\alpha]$ is the $S$-matrix in the background, while $S[0]$ is the usual vacuum $S$ matrix.
Assuming that the Vaidya metric in quantum field theory can indeed be sourced by the coherent state $\ket\alpha$, then we may relate the Vaidya vacuum state $\ket{\Omega}$ above to the coherent state:
\[
\ket{\Omega} = \ket{\alpha} \,.
\]
Now in the far past we just have flat space states. 
Given the properties of the displacement operator, we may either absorb the coherent state into the background (as we did) or work with it explicitly (to make more contact with flat space amplitudes).

\begin{figure}[t!]
\begin{center}
\tikzset{every picture/.style={line width=0.75pt}} 
\begin{tikzpicture}[x=0.9pt,y=0.9pt,yscale=-1,xscale=1]

\draw    (531.71,17502.68) -- (455.33,17489.52) ;
\draw [shift={(493.52,17496.1)}, rotate = 189.78] [fill={rgb, 255:red, 0; green, 0; blue, 0 }  ][line width=0.08]  [draw opacity=0] (5.36,-2.57) -- (0,0) -- (5.36,2.57) -- cycle    ;
\draw    (565.43,17492.97) -- (531.71,17502.68) ;
\draw [shift={(548.57,17497.83)}, rotate = 163.93] [fill={rgb, 255:red, 0; green, 0; blue, 0 }  ][line width=0.08]  [draw opacity=0] (5.36,-2.57) -- (0,0) -- (5.36,2.57) -- cycle    ;
\draw [color={rgb, 255:red, 0; green, 0; blue, 0 }  ,draw opacity=1 ]   (533.21,17502.69) .. controls (534.87,17504.37) and (534.86,17506.04) .. (533.18,17507.69) .. controls (531.5,17509.34) and (531.49,17511.01) .. (533.14,17512.69) .. controls (534.79,17514.37) and (534.78,17516.04) .. (533.1,17517.69) .. controls (531.42,17519.34) and (531.41,17521.01) .. (533.06,17522.69) .. controls (534.71,17524.37) and (534.7,17526.04) .. (533.02,17527.69) .. controls (531.34,17529.34) and (531.33,17531.01) .. (532.98,17532.69) .. controls (534.63,17534.36) and (534.62,17536.03) .. (532.95,17537.69) .. controls (531.27,17539.34) and (531.26,17541.01) .. (532.91,17542.69) .. controls (534.56,17544.37) and (534.55,17546.04) .. (532.87,17547.69) .. controls (531.19,17549.34) and (531.18,17551.01) .. (532.83,17552.69) .. controls (534.48,17554.37) and (534.47,17556.04) .. (532.79,17557.69) -- (532.79,17558.68) -- (532.79,17558.68)(530.21,17502.67) .. controls (531.87,17504.35) and (531.86,17506.02) .. (530.18,17507.67) .. controls (528.5,17509.32) and (528.49,17510.99) .. (530.14,17512.67) .. controls (531.79,17514.35) and (531.78,17516.02) .. (530.1,17517.67) .. controls (528.42,17519.32) and (528.41,17520.99) .. (530.06,17522.67) .. controls (531.71,17524.35) and (531.7,17526.02) .. (530.02,17527.67) .. controls (528.34,17529.32) and (528.33,17530.99) .. (529.98,17532.67) .. controls (531.63,17534.34) and (531.62,17536.01) .. (529.95,17537.67) .. controls (528.27,17539.32) and (528.26,17540.99) .. (529.91,17542.67) .. controls (531.56,17544.35) and (531.55,17546.02) .. (529.87,17547.67) .. controls (528.19,17549.32) and (528.18,17550.99) .. (529.83,17552.67) .. controls (531.48,17554.35) and (531.47,17556.02) .. (529.79,17557.67) -- (529.79,17558.66) -- (529.79,17558.66) ;
\draw    (538.43,17524.95) -- (538.43,17537.95) ;
\draw [shift={(538.43,17521.95)}, rotate = 90] [fill={rgb, 255:red, 0; green, 0; blue, 0 }  ][line width=0.08]  [draw opacity=0] (5.36,-2.57) -- (0,0) -- (5.36,2.57) -- cycle    ;
\draw    (567.71,17561.33) -- (489.99,17560.88)(567.72,17558.33) -- (490.01,17557.88) ;
\draw [shift={(528.86,17559.6)}, rotate = 180.33] [fill={rgb, 255:red, 0; green, 0; blue, 0 }  ][line width=0.08]  [draw opacity=0] (7.14,-3.43) -- (0,0) -- (7.14,3.43) -- cycle    ;
\draw [color={rgb, 255:red, 0; green, 0; blue, 0 }  ,draw opacity=1 ] [dash pattern={on 4.5pt off 4.5pt}]  (443.2,17534.58) -- (474.55,17559.3) ;
\draw [color={rgb, 255:red, 0; green, 0; blue, 0 }  ,draw opacity=1 ] [dash pattern={on 4.5pt off 4.5pt}]  (438.8,17542.98) -- (474.55,17559.3) ;
\draw [color={rgb, 255:red, 0; green, 0; blue, 0 }  ,draw opacity=1 ] [dash pattern={on 4.5pt off 4.5pt}]  (436,17552.98) -- (474.55,17559.3) ;
\draw [color={rgb, 255:red, 0; green, 0; blue, 0 }  ,draw opacity=1 ] [dash pattern={on 4.5pt off 4.5pt}]  (442.4,17582.88) -- (474.55,17559.3) ;
\draw  [fill={rgb, 255:red, 74; green, 144; blue, 226 }  ,fill opacity=1 ] (474.64,17543.86) .. controls (483.17,17543.91) and (490.05,17550.85) .. (490,17559.38) .. controls (489.95,17567.91) and (482.99,17574.79) .. (474.46,17574.75) .. controls (465.93,17574.7) and (459.05,17567.75) .. (459.1,17559.23) .. controls (459.15,17550.7) and (466.11,17543.82) .. (474.64,17543.86) -- cycle ;
\draw  [draw opacity=0][dash pattern={on 0.84pt off 2.51pt}] (441.86,17576.35) .. controls (439.68,17573.58) and (438.11,17569.83) .. (437.57,17565.58) .. controls (437.14,17562.16) and (437.44,17558.88) .. (438.32,17556.05) -- (449.62,17564.06) -- cycle ; \draw  [dash pattern={on 0.84pt off 2.51pt}] (441.86,17576.35) .. controls (439.68,17573.58) and (438.11,17569.83) .. (437.57,17565.58) .. controls (437.14,17562.16) and (437.44,17558.88) .. (438.32,17556.05) ;  

\draw (570.71,17485.95) node [anchor=north west][inner sep=0.75pt]    {$p'$};
\draw (441.43,17487.05) node [anchor=north west][inner sep=0.75pt]    {$p$};
\draw (543.29,17526.81) node [anchor=north west][inner sep=0.75pt]    {$q$};
\draw (455.31,17527.3) node [anchor=north west][inner sep=0.75pt]  [font=\small,xslant=0.03]  {$H_{\alpha \rightarrow \text{BH}}$};
\draw (490.86,17565.81) node [anchor=north west][inner sep=0.75pt]    {$P+q$};
\draw (573.29,17559.95) node [anchor=north west][inner sep=0.75pt]    {$P$};
\draw (413.33,17556.68) node [anchor=north west][inner sep=0.75pt]    {$\alpha $};
\end{tikzpicture} 
\caption{Hawking scattering as 
a  dynamical process. The form factor ${H}_{\alpha\to \text{BH}}$ describes the collapse of the coherent state $\alpha$ into a black hole with momentum $P+q$. }\label{bhqft1}
\end{center}
\end{figure}
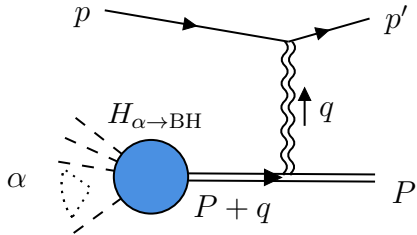

However in order to perform a concrete computation we need an additional ingredient: our infalling matter must transition into a black hole.
The Vaidya metric clearly incorporates this ingredient above.
It is obviously hopeless to fully track the detailed quantum gravitational details of the black hole formation process; however, these details are also not necessary to reproduce our work in flat space field theory.
All that is necessary is that a black hole is formed.
This could perhaps be achieved by introducing some form factor $H_{\alpha \rightarrow \textrm{BH}}$ for the transition (and perhaps some inclusive sum) along the lines used in hadronic physics.
A Feynman-style diagram for the Hawking scattering process, in flat space, is shown in figure~\ref{bhqft1}.
It is sufficient to study single graviton exchange between the infalling radiation/black hole system and the scattering particle, because we know from section~\ref{sec:Tree_level} that the full scattering can be understood by exponentiating these diagrams (to leading order in the phase).
After exponentiating, the pair production process is recovered by crossing.
There is no need to worry about producing many pairs because we understand the nature of the squeezing process; this is analogous to studying gravitational wave processes by computing single-graviton production amplitudes.

We hope that some of these ideas will be a beginning for more concrete work on black hole greybody factors, the backreaction of Hawking radiation on the black hole system, perhaps the Page curve, among other exciting topics.

\acknowledgments
We thank Simon Caron-Huot, Andrea Cristofoli, Hofie Hannesd\'ottir, Sebastian Mizera, Julio Parra-Martinez, Gui Pimentel, Grant Remmen, Radu Roiban, Alex Ochirov, L\'arus Thorlacius and Fei Teng for discussions and useful comments.
We thank Andrea, Radu, Grant, Gui and Simon, and especially Alex, for their comments on an early draft of this paper. 
Anton Ilderton has been particularly helpful over an extended period of time.
M.S.  has been supported by the European Research Council under Advanced Investigator grant [ERC–AdG–885414]. R.A. and DOC thank David Kosower and  the financial support from  the grant 
[ERC –AdG–885414] during their visit to the IPhT in the spring of 2024.
R.A. is supported by UK Research and Innovation (UKRI) under the UK
government’s Horizon Europe Marie Skłodowska-Curie funding guarantee grant [EP/Z000947/1] and by the STFC grant “Particle Theory at the Higgs Centre”.
DOC thanks for the IAS for hospitality, and is supported by the STFC grant “Particle Theory at the Higgs Centre” and by the European Research Council under Advanced Investigator grant
ERC–AdG–101200505.
We also thank the authors of reference~\cite{Copinger:2024pai} for sharing a draft of their paper prior to publication.
Some manipulations were performed with the help of \texttt{FeynCalc}~\cite{MERTIG1991345,Shtabovenko:2016sxi,Shtabovenko:2020gxv}. 
The Penrose diagram was made with TikZ~\cite{Ellis:2016jkw}.

\appendix

\section{Spherical States and Wavefunctions}
\label{app:sphericalNorms}

To take advantage of the spherical symmetry of the Vaidya metric, it is sometimes convenient for us to use spherically symmetric states. 
While plane wave states diagonalize $P^\mu$, partial wave states are eigenstates of $\bm{J}^2$, $J_z$ and $P^0$.
We normalise our (massless scalar) states without introducing any additional dimensionful factors with respect to the usual plane wave states in order to preserve the familiar dimensional analysis of scattering amplitudes.
We define therefore define partial wave states as
\[
\ket{p \, \mathcal{\ell} \, m} = \frac{1}{\sqrt{4\pi}} \int \dd \Omega_{p} \, Y_{\ell m}(\Omega_p) \ket{p} \,,
\]
where $Y_{\ell m}$ are the spherical harmonics and the integration is over the angles of the spatial vector components of $p^\mu$ in the privileged frame of the metric.
We are mostly interested in the spherically symmetric $\ell = 0$, $m=0$ state which is then
\[
\ket{p \, 0 \, 0} = \frac{1}{4\pi} \int \dd \Omega_{p}  \ket{p} \,.
\]
(To be clear, we use the spherical harmonic conventions of DLMF \cite{NIST:DLMF}.)

We only require the wavefunction for a spherical $\ell = 0$, $m=0$ state of energy $E$ which is
\[ \label{eq:sphericalWFnorm}
\braket{\vac | \phi(x) | E \, 0 \, 0} &= \frac{1}{4\pi} \int \dd \Omega_p \, e^{-i E t} e^{i E r \cos \theta} \\
&= \frac{-i}{2 E r} \left ( e^{-iE u} - e^{-i E v} \right) \,.  
\]
Note that $u = t-r$ and $v = t + r$, so the two terms in the last equality above correspond to the outgoing and incoming waves respectively.

In terms of plane wave states, the single-particle completeness relation reads
\[
1 = \int \dd \Phi(p) \ket{p} \bra{p} = \frac{1}{16 \pi^3} \int \dd p \, p \int \dd \Omega_p \ket{p} \bra{p} \,.
\]
Taking advantage of the completeness of spherical harmonics,
\[
\sum_{\ell, m} Y^*_{\ell m} (\theta', \phi') Y_{\ell m} (\theta, \phi) = \delta(\cos \theta - \cos \theta') \delta(\phi - \phi') \,,
\]
we can rewrite the single-particle completeness relation as
\[\label{eq:completenessELM}
1 = \sum_{\ell, m} \int \dd E \frac{E}{4 \pi^2}  \ket{E \, \ell \, m}\bra{E \, \ell \, m} \,.
\]
The factor $E / 4 \pi^2$ appearing here is relevant for computing the number distribution of emitted spherical quanta in the main text.

Other normalisations of plane wave states are very common in the literature. 
In particular, many authors normalise their partial-wave states to remove the factor $E$ from the completeness relation~\eqref{eq:completenessELM}. 
This has the advantage that the states are more simply normalised, and the disadvantage that square roots of energy appear in various other places (notably in wavefunctions).
The conventions outlined above were convenient for us because the relation between plane wave states and (especially) spherical $\ell = 0$, $m=0$ states is particularly simple.

\bibliographystyle{JHEP}
\bibliography{HawkingAmplitudes.bib}
\end{document}